\documentclass[english,preprint,floatfix]{revtex4}
\usepackage[T1]{fontenc}
\usepackage[latin1]{inputenc}
\usepackage{amsmath}
\usepackage{graphicx}
\usepackage{amssymb}

\makeatletter
\usepackage{babel}
\makeatother
\begin{document}

\title{Microscopic calculation of $^{240}\textrm{Pu}$ scission with a finite-range
effective force}

\author{W. Younes, and D. Gogny}

\address{Lawrence Livermore National Laboratory, Livermore, CA 94551}

\date{\today}

\begin{abstract}
Hartree-Fock-Bogoliubov calculations of hot fission in $^{240}\textrm{Pu}$
have been performed with a newly-implemented code that uses the D1S
finite-range effective interaction. The hot-scission line is identified
in the quadrupole-octupole-moment coordinate space. Fission-fragment
shapes are extracted from the calculations. A benchmark calculation
for $^{226}\textrm{Th}$ is obtained and compared to results in the
literature. In addition, technical aspects of the use of HFB calculations
for fission studies are examined in detail. In particular, the identification
of scission configurations, the sensitivity of near-scission calculations
to the choice of collective coordinates in the HFB iterations, and
the formalism for the adjustment of collective-variable constraints
are discussed. The power of the constraint-adjustment algorithm is
illustrated with calculations near the critical scission configurations
with up to seven simultaneous constraints.
\end{abstract}
\maketitle

\section{Introduction}

The last three decades have seen a resurgence of interest in the microscopic
description of nuclear fission. This renaissance in fission theory
has been ushered in by progress in formal many-body theory and by
the advent of faster and parallel computers. The microscopic approach
can boast a well-established track record of accomplishment over the
last three decades, such as the prediction of fission barriers \cite{berger84,delaroche06,flocard74,warda02,perez-martin09,rutz95,burnevich04},
and their evolution with temperature \cite{martin09} and angular
momentum \cite{egido00}, the prediction of fission times \cite{berger84,smolanczuk95}
and fission-isomer lifetimes \cite{chinn92}, the description of hot
and cold fission \cite{berger84}, the prediction of fission yields
\cite{goutte05}, the description of cluster radioactivity as very
asymmetric fission \cite{robledo08}, and most recently, the calculation
of fission-fragment properties (e.g., excitation energy, shape, kinetic
energy, emitted-neutron multiplicity, angular momentum) \cite{dubray08,bonneau07}.
Despite these successes however, the microscopic description of fission
remains one of the most difficult challenges in nuclear physics.

On the other hand, the promise of a microscopic theory that can reliably
predict nearly all aspects of fission within a single, self-consistent
framework is tantalizing. A fully self-consistent, dynamical approach
to fission has been developed by the group at Bruyères-le-Châtel \cite{berger84,goutte05,dubray08},
and is being implemented at Livermore \cite{younes07}. This approach
treats both static and dynamic aspects of fission self-consistently
and requires as its only phenomenological input the effective interaction
between the nucleons.

A Hartree-Fock-Bogoliubov (HFB) code is the central tool for the description
of the static aspects of fission in the microscopic method. The use
of a finite-range effective interaction, such as the D1S interaction
\cite{gogny75}, allows for the treatment of pairing within the HFB
formalism \cite{bogoliubov58} in a fully self-consistent manner,
and without the need for additional phenomenological parameters. The
HFB calculations can be constrained by a judicious choice of collective
variables to explore those nuclear shapes that are relevant to fission.
Such constraints have confirmed the richness of fission phenomena,
for example by revealing the full range of fission modes from hot
(fragments formed in maximally-excited states) to cold (fragments
formed with no excitation energy) \cite{berger84}.

In the dynamical component of the microscopic theory, a wave packet
is built from HFB solutions constrained over all relevant nuclear
shapes using the Time-Dependent Generator-Coordinate Method (TDGCM)
\cite{hill53,griffin57,haider92,reinhard83}. In practical applications,
the Gaussian-Overlap Approximation (GOA) to the TDGCM can be used
to produce a collective Schrödinger equation, and therefore a collective
Hamiltonian, constructed entirely from the single-particle degrees
of freedom. The TDGCM formalism describes the nucleus in its lowest-energy
state, as well as its collective excitations \cite{onishi70,bonche90},
and can be extended to include intrinsic excitations as well \cite{didong76}
on the way to scission. These intrinsic excitations are needed for
a microscopic description of fission that goes beyond the standard
adiabatic approximation usually adopted in fission calculations \cite{schutte81}.
This comprehensive program for the microscopic description of induced
fission has already shown the importance of dynamical effects in the
prediction of fission times \cite{berger84} and fission-fragment
yields \cite{goutte05}, but a great deal of work remains to include
all the relevant physics aspects in the calculation. In particular,
a detailed and quantitative understanding of scission itself remains
to be developed even at the level of the static calculations.

In this paper, we focus on the static aspect of the microscopic theory
with three goals in mind: 1) to introduce the newly-developed HFB
code FRANCHBRIE \cite{younes07}, which uses a finite-range effective
interaction, 2) to examine in detail some basic technical aspects
of fission calculations with an HFB code, and 3) to present first-time
results of scission properties for the hot fission of $^{240}\textrm{Pu}$.
In section \ref{sec:Theory} we review the HFB formalism and discuss
in detail some features of the one-center deformed harmonic-oscillator
basis, formal and practical aspects of HFB fission calculations with
multiple constraints, as well as the HFB convergence algorithm itself.
In section \ref{sec:Results}, we benchmark our HFB code against two-center
calculations of scission properties for $^{226}\textrm{Th}$ by Dubray
\emph{et al}. \cite{dubray08}. We then apply the code to the identification
of hot-scission configurations in $^{240}\textrm{Pu}$, and the shapes
of the nascent fragments just before scission.

\section{Theory\label{sec:Theory}}

\subsection{General HFB formalism\label{sub:General-HFB-formalism}}

For convenience, we recall the main points of the HFB formalism with
a finite-range effective interaction in this section and refer the
reader to the literature for further details (see, e.g., \cite{ring80,girod83,decharge80}).
We have implemented this formalism within the code FRANCHBRIE \cite{younes07}.

We start from the many-body Hamiltonian in second-quantized notation
(see, e.g., chapter 5 in \cite{ring80}),\begin{eqnarray*}
H & = & \sum_{mn}t_{mn}a_{m}^{\dagger}a_{n}+\frac{1}{4}\sum_{mnpq}\bar{\mathcal{V}}_{mnpq}a_{m}^{\dagger}a_{n}^{\dagger}a_{q}a_{p}\end{eqnarray*}
with the antisymmetrized two-body matrix elements\begin{eqnarray*}
\bar{\mathcal{V}}_{mnpq} & \equiv & \left\langle mn\left|\mathcal{V}\right|pq\right\rangle -\left\langle mn\left|\mathcal{V}\right|qp\right\rangle \end{eqnarray*}
and the usual anticommutation rules for particle operators\begin{equation}
\left\{ a_{m},a_{n}\right\} =\left\{ a_{m}^{\dagger},a_{n}^{\dagger}\right\} =0,\quad\left\{ a_{m}^{\dagger},a_{n}\right\} =\delta_{mn}\label{eq:particle-anticomm}\end{equation}
In this paper, we use a finite-range effective interaction which in
coordinate space takes the form \cite{decharge80}\begin{eqnarray}
 &  & \mathcal{V}\left(\vec{r}_{1},\vec{r}_{2}\right)\nonumber \\
 & = & \sum_{i=1}^{2}\left(W_{i}+B_{i}\hat{P}_{\sigma}-H_{i}\hat{P}_{\tau}-M_{i}\hat{P}_{\sigma}\hat{P}_{\tau}\right)e^{-\left(\vec{r}_{1}-\vec{r}_{2}\right)^{2}/\mu_{i}^{2}}\nonumber \\
 &  & +iW_{LS}\overleftarrow{\nabla}_{12}\times\delta\left(\vec{r}_{1}-\vec{r}_{2}\right)\overrightarrow{\nabla}_{12}\cdot\left(\vec{\sigma}_{1}+\vec{\sigma}_{2}\right)\nonumber \\
 &  & +t_{0}\left(1+x_{0}\hat{P}_{\sigma}\right)\delta\left(\vec{r}_{1}-\vec{r}_{2}\right)\rho^{\gamma}\left(\frac{\vec{r}_{1}+\vec{r}_{2}}{2}\right)+V_{\textrm{Coul}}\label{eq:gogny-int}\end{eqnarray}
where $\overleftarrow{\nabla}_{12}\equiv\overleftarrow{\nabla}_{1}-\overleftarrow{\nabla}_{2}$
, $\overrightarrow{\nabla}_{12}\equiv\overrightarrow{\nabla}_{1}-\overrightarrow{\nabla}_{2}$,
$\hat{P}_{\sigma}$ is the spin-exchange operator, and $\hat{P}_{\tau}$
is the isospin-exchange operator. The Coulomb interaction $V_{\textrm{Coul}}$
is added if both particles are protons, and $\rho\left(\vec{r}\right)$
denotes the total nuclear density. The D1S effective interaction \cite{berger84,warda02}
has been used for the present calculations. Given the computationally-intensive
nature of the calculations, we have omitted contributions from the
spin-orbit and Coulomb interactions to the pairing field. This approximation
is well justified in the case of the spin-orbit interaction whose
intensity in the singlet-even channel is very weak, but less so for
the Coulomb term that can significantly reduce the pairing correlations
for proton pairs \cite{anguiano01}. We note also that the density-dependent
part of the interaction is adjusted to cancel in the singlet-even
channel by setting $x_{0}=1$. Consequently, only the Gaussian terms
contribute to the pairing field, which permits the fully self-consistent
application of the Bogoliubov formalism, without the need for arbitrary
truncations of the space or the use of ad-hoc pairing forces. The
Coulomb exchange contribution has been treated in the Slater approximation,
and the two-body contribution to the center-of-mass correction has
been included in the mean field.

The Bogoliubov theory \cite{bogoliubov58} takes into account, in
an approximate way, two-body correlations beyond the mean-field restriction
to particle-hole excitations. The approach defines quasiparticle creation
and destruction operators as linear combinations of the particle creation
and destruction operators,\begin{eqnarray}
\eta_{\mu}^{\dagger} & \equiv & \sum_{n}\left(U_{n\mu}a_{n}^{\dagger}+V_{n\mu}a_{n}\right)\nonumber \\
\eta_{\mu} & \equiv & \sum_{n}\left(U_{n\mu}^{*}a_{n}+V_{n\mu}^{*}a_{n}^{\dagger}\right)\label{eq:bogo-trans}\end{eqnarray}
Assuming there exists a vacuum of the destruction operators $\eta_{\mu}$,
denoted by $\left|\tilde{0}\right\rangle $, we identify it as the
ground state of the nucleus and its energy can be written simply as
a functional of the density matrix and the pairing tensor or, equivalently,
as a functional of the generalized density\begin{eqnarray}
R & \equiv & \left(\begin{array}{cc}
\rho & -\kappa\\
\kappa^{*} & I-\rho^{*}\end{array}\right)\equiv\left(\begin{array}{cc}
R^{11} & R^{12}\\
R^{21} & R^{22}\end{array}\right)\label{eq:R-def}\end{eqnarray}
We recall that the unitarity condition of the transformation in Eq.
(3) is equivalent to\begin{eqnarray}
R^{2} & = & R\label{eq:R-idempotence}\end{eqnarray}
we will therefore write the energy as\begin{eqnarray}
 &  & E\left(\rho,\kappa,\lambda_{p},\lambda_{n},\Lambda\right)\nonumber \\
 & = & E\left(\rho,\kappa\right)-\lambda_{p}\left\langle \tilde{0}\left|\hat{N}_{p}\right|\tilde{0}\right\rangle -\lambda_{n}\left\langle \tilde{0}\left|\hat{N}_{n}\right|\tilde{0}\right\rangle \nonumber \\
 &  & -\textrm{Tr}\left[\Lambda\left(R^{2}-R\right)\right]\label{eq:energy-functional1}\end{eqnarray}
where $E\left(\rho,\kappa\right)$ is the expectation value of the
Hamiltonian in the quasiparticle ground state, $\lambda_{p}$ and
$\lambda_{n}$ are the Lagrange parameters needed to impose the appropriate
average number of protons and neutrons, respectively, given by the
matrix $R$. The matrix $\Lambda$ of Lagrange parameters is needed
to satisfy Eq. (\ref{eq:R-idempotence}). Thus the determination of
the fundamental nuclear state amounts to finding the generalized density
matrix that minimizes Eq. (\ref{eq:energy-functional1}). Some authors
recognize Eq. (\ref{eq:energy-functional1}) as the equation of a
multidimensional surface, and seek its minimum directly using standard
mathematical techniques to find the minimum of a function. Among these
approaches, we cite the gradient method \cite{egido80} or an improved
variant known as the conjugate gradient method \cite{egido95}. The
number and diversity of applications using this method speak to its
effectiveness \cite{robledo08a,aunguiano02,perez-martin09,martin09,egido00,robledo08}.
In our approach to the minimization of Eq. (\ref{eq:energy-functional1}),
we start with the variational principle,\begin{eqnarray}
\delta E\left(\rho,\kappa,\lambda_{p},\lambda_{n},\Lambda\right) & = & \textrm{Tr}\left\{ \left[\mathcal{H}-\left(\Lambda R+R\Lambda-\Lambda\right)\right]\delta R\right\} \nonumber \\
 & = & 0\label{eq:energy-variation}\end{eqnarray}
$\forall\delta R$ where\begin{eqnarray}
\mathcal{H}_{mn}^{ij} & \equiv & 2\frac{\delta E\left(\rho,\kappa,\lambda_{p},\lambda_{n}\right)}{\delta R_{nm}^{ji}}\label{eq:genH-def}\end{eqnarray}
Taking into account Eq. (\ref{eq:R-idempotence}) it is possible to
eliminate the constraint matrix $\Lambda$, leading to the Bogoliubov
equation\begin{eqnarray}
\left[\mathcal{H}\left(R\right),R\right] & = & 0\label{eq:HFB-equation}\end{eqnarray}
The Bogoliubov matrix $\mathcal{H}$ in Eq. (\ref{eq:HFB-equation})
is constructed with the help of the block matrices defined by Eq.
(\ref{eq:genH-def}). The explicit form of these matrix elements for
the D1S effective interaction is given by references \cite{decharge80,younes09}.
The solution of Eq. (\ref{eq:HFB-equation}) is then found by successive
diagonalizations of the Bogoliubov Hamiltonian. This iterative solution
method is described in greater detail in section \ref{sub:The-HFB-convergence}
and appendix \ref{sec:Multiple-constraint-formalism}.

\subsection{Basis truncation and aspects of one-center basis calculations\label{sub:Basis-truncation-and}}

In practical applications, the formalism of section \ref{sub:General-HFB-formalism}
must be expressed in some basis. Typically, the deformed Harmonic-oscillator
(HO) basis (see, e.g., chapter 2 in \cite{ring80}) has been used
in many HFB calculations, including those dealing with fission \cite{bonneau06,warda02}.
The basis states in cylindrical coordinates $\left(\rho,z,\varphi\right)$
are\begin{eqnarray}
\left\langle \left.\vec{r}\right|n_{r},\Lambda,n_{z},\sigma\right\rangle  & = & \Phi_{n_{r},\left|\Lambda\right|}\left(\rho;b_{\perp}\right)\frac{e^{\mathrm{i}\Lambda\varphi}}{\sqrt{2\pi}}\nonumber \\
 &  & \times\Phi_{n_{z}}\left(z;b_{z}\right)\chi_{\sigma}\label{eq:ho-fun}\end{eqnarray}
where the explicit forms used in this work for the radial ($\Phi_{n_{r},\left|\Lambda\right|}$)
and Cartesian ($\Phi_{n_{z}}$) components and their relevant properties
can be found, e.g., in \cite{younes09}, and $\chi_{\sigma}$ is a
spinor function for $\sigma=\pm1/2$. These basis states assume axial
symmetry of the nucleus explicitly. Other symmetries can also be imposed
on the HFB calculation to reduce the overall size of the problem.
Two symmetries in particular are relevant to the fission calculations
in this paper: the symmetry with respect to the parity operator $\hat{\Pi}$\begin{eqnarray*}
\hat{\Pi}\left|n_{r},\Lambda,n_{z},\sigma\right\rangle  & = & (-1)^{\left|\Lambda\right|+n_{z}}\left|n_{r},\Lambda,n_{z},\sigma\right\rangle \end{eqnarray*}
and the symmetry with respect to the z-signature operator $\hat{S}_{z}=i\hat{R}_{z}\left(\pi\right)$,
where $\hat{R}_{z}\left(\pi\right)$ effects a rotation by $\pi$
in both spatial and spin space,

\begin{eqnarray*}
\hat{S}_{z}\left|n_{r},\Lambda,n_{z},\sigma\right\rangle  & = & \sigma(-1)^{\left|\Lambda\right|}\left|n_{r},\Lambda,n_{z},\sigma\right\rangle \end{eqnarray*}
Throughout this work, only the z-signature symmetry has been imposed,
leaving the fissioning nucleus free to violate the symmetry with respect
to parity and assume asymmetric shapes. These symmetries are taken
into account explicitly by rewriting the general Bogoliubov transformation
of Eq. (\ref{eq:bogo-trans}) in terms of the relevant quantum numbers
as\begin{eqnarray*}
\eta_{\mu}^{\dagger}\left(q,s_{z},\Omega\right) & \equiv & \sum_{n}\left[U_{n\mu}^{q,s_{z},\Omega}a_{n}^{\dagger}\left(q,s_{z},\Omega\right)\right.\\
 &  & \left.+V_{n\mu}^{q,s_{z},\Omega}a_{n}\left(q,s_{z},\bar{\Omega}\right)\right]\\
\eta_{\mu}\left(q,s_{z},\bar{\Omega}\right) & \equiv & \sum_{n}\left[\left(U_{n\mu}^{q,s_{z},\bar{\Omega}}\right)^{*}a_{n}\left(q,s_{z},\bar{\Omega}\right)\right.\\
 &  & \left.+\left(V_{n\mu}^{q,s_{z},\bar{\Omega}}\right)^{*}a_{n}^{\dagger}\left(q,s_{z},\Omega\right)\right]\end{eqnarray*}
where $q$ distinguishes protons and neutrons, $s_{z}=\pm1$ is the
z-signature quantum number, and $\bar{\Omega}$ is the total angular-momentum
projection for the time-reversed state.

Even with the z-signature symmetry imposed, the treatment of fission
can require large basis sizes and the calculation of a large number
of two-body matrix elements. In order to further limit the size of
problem, various basis truncation schemes have been devised. Some
\cite{flocard73} keep only those basis states with corresponding
HO energies below a given cutoff, while other schemes \cite{egido97,warda02}
directly allow for more quanta along the $z$ direction--the direction
of elongation of the fissioning nucleus--compared to the radial direction.
In the truncation scheme of \cite{flocard73}, the HO quantum numbers
must satisfy\begin{eqnarray}
\hbar\omega_{\bot}\left(n_{\bot}+1\right)+\hbar\omega_{z}\left(n_{z}+\frac{1}{2}\right) & \leq & \hbar\omega_{0}\left(N+2\right)\label{eq:flocard-trunc}\end{eqnarray}
with $n_{\bot}\equiv2n_{r}+\left|\Lambda\right|$ and for a given
maximum shell number $N$, where the oscillator frequencies are related
to the length parameters $b_{\perp}$ and $b_{z}$ in Eq. (\ref{eq:ho-fun})
by\begin{equation}
\omega_{\bot}=\frac{\hbar}{mb_{\perp}^{2}},\qquad\omega_{z}=\frac{\hbar}{mb_{z}^{2}},\qquad\omega_{0}^{3}=\omega_{\bot}^{2}\omega_{z}\label{eq:ho-freq-def}\end{equation}
and $m$ is the nucleon mass. With increasing axial elongation and
for fixed $N$, Eq. (\ref{eq:flocard-trunc}) adds more shells in
the $z$ direction while simultaneously decreasing the number of shells
in the radial direction, thus keeping the basis size from growing
too quickly with deformation. In the truncation scheme of \cite{egido97,warda02},
the condition\begin{eqnarray}
\frac{n_{z}}{q}+2n_{r}+\left|\Lambda\right| & \leq & N\label{eq:egido-trunc}\end{eqnarray}
is imposed for a given maximum shell number $N$ and parameter $q$.
In this work we have used both truncation schemes. The truncation
given by Eq. (\ref{eq:flocard-trunc}) has been used for most calculations
in this paper, while the truncation of Eq. (\ref{eq:egido-trunc})
has been used mainly in section \ref{sub:Benchmark:226Th}.

The oscillator lengths $b_{\perp}$ and $b_{z}$ in Eq. (\ref{eq:ho-fun}),
or equivalently the frequencies $\omega_{\bot}$ and $\omega_{z}$,
are variational parameters in the HFB calculation that must be chosen
to minimize the HFB energy. Through a series of calculations in $^{240}\textrm{Pu}$
using the truncation scheme of Eq. (\ref{eq:egido-trunc}) with $N=13$
and $q=1.5$, and exploring a wide range of values of the constraints
on the quadrupole ($Q_{20}$) and octupole ($Q_{30}$) moments, an
approximate dependence was obtained for the frequencies that minimize
the HFB energy, given by\begin{eqnarray}
\hbar\omega_{0} & = & 8.4345-0.0021668\, Q_{20}\label{eq:parm-hw0}\\
\frac{\omega_{\bot}}{\omega_{z}} & = & 1.7041+0.0028743\, Q_{20}\label{eq:parm-q}\end{eqnarray}
with $Q_{20}$ in barns and $\hbar\omega_{0}$ in MeV. No significant
dependence on $Q_{30}$ was observed in the range of interest.

Perhaps the most important aspect of the basis states in Eq. (\ref{eq:ho-fun})
is that they are centered about the origin by construction. In particular,
the Gaussian factor in Eq. (\ref{eq:ho-fun}) ensures that the nuclear
wave function falls off rapidly with increasing $z$. Despite this
feature of the basis states, we will show that it is still possible
to describe the exotic shapes occurring in fission. In order to describe
both the neck (near $z=0$) and nascent fragments (typically 5-10
fm from the origin) with the basis states of Eq. (\ref{eq:ho-fun}),
we are forced to include many quanta in the $z$ direction, and to
use relatively large values of $b_{z}$.

\begin{figure}
\includegraphics[%
  scale=0.33,
  angle=-90]{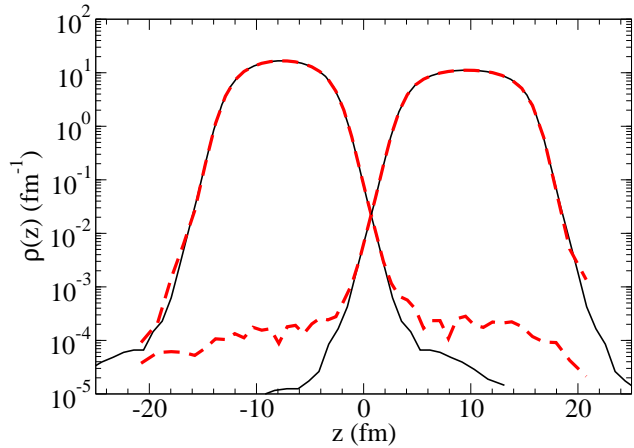}

\caption{\label{cap:transfrag}(Color online) Plots of the nuclear densities
for fragments of $^{134}\textrm{Te}$ and $^{106}\textrm{Mo}$ along
the axis of elongation of the nucleus, calculated in the one-center
basis and plotted (as solid black lines) centered at z = -7.63 and
9.65 fm, respectively. The dashed red lines represent the same densities,
but translated from the origin to their respective centroid positions
within a finite harmonic-oscillator-basis truncated according to Eq.
(\ref{eq:flocard-trunc}) and with $N$=13 shells, using the formalism
in appendix \ref{sec:Translation-in-a}.}
\end{figure}

\begin{figure}
\includegraphics[%
  scale=0.33,
  angle=-90]{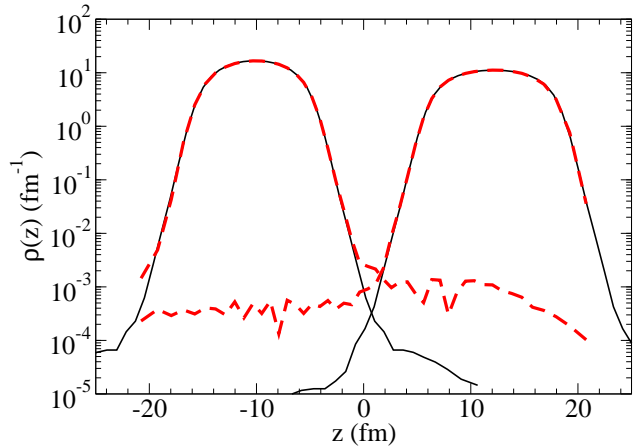}

\caption{\label{cap:transfrag1}(Color online) Same as Fig. \ref{cap:transfrag},
but for the $^{134}\textrm{Te}$ and $^{106}\textrm{Mo}$ fragments
translated an additional 2.5 fm each, to centroids at z = -10.13 and
12.15 fm, respectively.}
\end{figure}

To justify the use of the one-center basis for the range of fissioning
configurations and quantities examined in this paper, we have performed
separate HFB calculations for $^{134}\textrm{Te}$ and $^{106}\textrm{Mo}$
centered at the origin, and translated the resulting wave functions
to the typical positions these nuclei occupy as $^{240}\textrm{Pu}$
nascent fission fragments. The formalism required for translating
a wave function expressed within a finite HO basis is given in appendix
\ref{sec:Translation-in-a}. The basis was truncated according to
Eq. (\ref{eq:flocard-trunc}) with $N=13$, resulting in a maximum
number $n_{z}=26$ along the $z$ axis. The result is shown in Fig.
\ref{cap:transfrag}, and compared to a translation in an infinite-sized
basis (obtained in practice by redrawing the curves at the displaced
centroid positions while preserving their shape). The comparison clearly
shows the appearance of spurious tails for each fragment translated
within a finite-size basis. If the fragments are separated further,
e.g. by an additional 2.5 fm for each fragment in Fig. \ref{cap:transfrag1},
the tails grow larger. However, the tails caused by the translation
in a finite basis remain relatively small ($\sim10^{-4}\;\textrm{fm}^{-1}$
in Fig. \ref{cap:transfrag}, and $\sim5\times10^{-4}\;\textrm{fm}^{-1}$
in Fig. \ref{cap:transfrag1}), and the separations between the fragments
in both figures are larger than those encountered in the remainder
of this work. In section \ref{sub:Benchmark:226Th} we will show that
these tails do not significantly affect the nuclear properties calculated
in this paper. In a forthcoming publication \cite{younes09a} we will
explore a more microscopic definition of scission and of the fission
fragments, and we will calculate quantities such as the interaction
energy between the fragments that may be more sensitively affected
by the presence of these tails \cite{younes09b}.

\subsection{Multiple constraints in HFB calculations\label{sub:Multiple-constraints-in}}

In this section, we focus on formal and practical considerations in
the choice and control of multiple constraints in HFB calculations.
We will describe a mechanism for the adjustment of the constraints
which generalizes the discussion in \cite{decharge80}. The formalism
described here and used in our calculations is that of variation with
linear constraints. Other approaches for the adjustment of constraints,
such as the quadratic-constraint method can also be found in the literature
\cite{giraud70}. We have adopted the linear-variation approach in
our work because we have found it to be stable and robust, and these
are important qualities needed to map out the scission configurations,
which requires precise control of the nuclear shape. For a process
like fission, these constraints are central not only to being able
to drive the nucleus to scission, but also to uncover the full richness
of the microscopic method in its ability to describe the complexities
of fission. In section \ref{sub:General-HFB-formalism} we already
discussed the introduction of constraints on the average number of
neutrons and protons for the HFB Hamiltonian. Further constraints
can be introduced through the external-field one-body operators $\lambda_{i}\hat{F}_{i}$,\begin{equation}
H-\sum_{i}\lambda_{i}\hat{F}_{i}\label{eq:constrd-ham}\end{equation}
where the parameters $\lambda_{i}$ are used to adjust the field intensities.
Based on Eq. (\ref{eq:HFB-equation}), the Bogoliubov equation associated
with Eq. (\ref{eq:constrd-ham}) can now be written\begin{eqnarray*}
\left[\mathcal{H}\left(R\right)-\sum_{i}\lambda_{i}\mathbb{F}_{i},R\right] & = & 0\end{eqnarray*}
where\begin{eqnarray}
\mathbb{F}_{i} & \equiv & \left(\begin{array}{cc}
\hat{F}_{i} & 0\\
0 & -\hat{F}_{i}^{*}\end{array}\right)\label{eq:F-mat-struct}\end{eqnarray}
in the particle-hole representation, and $\mathcal{H}\left(R\right)$
is given by Eq. (\ref{eq:genH-def}). In what follows, we will use
the notation\begin{eqnarray*}
\mathcal{H}\left(R,\left\{ \lambda_{i}\right\} \right) & \equiv & \mathcal{H}\left(R\right)-\sum_{i}\lambda_{i}\mathbb{F}_{i}\end{eqnarray*}
where $\left\{ \lambda_{i}\right\} $ represents the set of Lagrange
multipliers other than those associated with the proton and neutron
numbers. The $\lambda_{i}$ Lagrange multipliers can be adjusted to
yield an HFB solution with desired expectation values $f_{i}$ of
the fields\begin{eqnarray*}
\left\langle \hat{F}_{i}\right\rangle  & = & \frac{1}{2}\textrm{Tr}\hat{F}_{i}+\frac{1}{2}\textrm{Tr}\mathbb{F}_{i}R\\
 & = & f_{i}\end{eqnarray*}
The formalism used to find the appropriate $\lambda_{i}$ parameters
is derived in appendix \ref{sec:Multiple-constraint-formalism}. In
describing fission within the microscopic approach, we are free to
impose any number of constraints, each defined by a corresponding
external-field operator. We are limited in this task by the computational
requirements, which grow quickly with the number of constraints, and
by their relevance to the fission process.

In the simplest physical picture of fission, we expect that the nucleus
will stretch along its symmetry axis until scission, and therefore
introduce the mass quadrupole operator $\hat{Q}_{20}$ as a constraint.
Next, the octupole operator $\hat{Q}_{30}$ is introduced to account
for the range of mass divisions observed in fragments, from symmetric
to asymmetric. With the introduction of the octupole constraint, we
are forced to impose a constraint on the dipole moment, $\hat{Q}_{10}$,
as well in order to maintain the center of mass of the nucleus fixed.
The hexadecapole operator $\hat{Q}_{40}$ controls the formation of
the neck between nascent fragments, and accounts for the range of
fission modes from cold to hot \cite{berger84}. In addition, we recall
that the HFB procedure requires constraints on the expected values
of the proton-number ($\hat{N}_{p}$) and neutron-number ($\hat{N}_{n}$)
operators. 

\begin{figure}
\includegraphics[%
  scale=0.33]{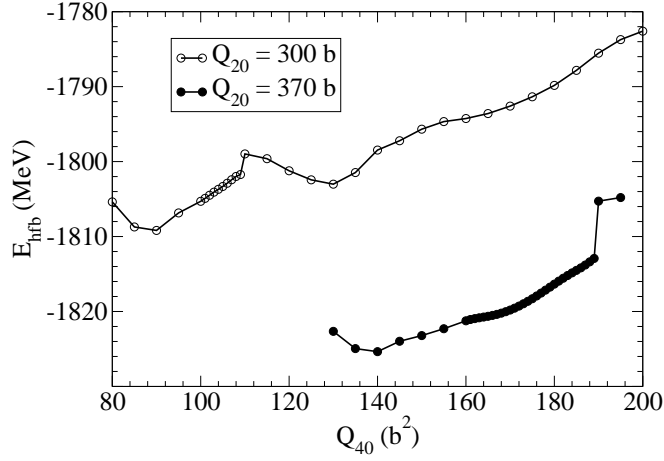}

\caption{\label{cap:plot-300-and-370-ehfb}Calculated HFB energy for $^{240}\textrm{Pu}$
as a function of hexadecapole moment, and for quadrupole moments of
300 b (cold fission) and 370 b (hot fission). For the $\left\langle Q_{20}\right\rangle =300\,\textrm{b}$
case, the fission valley is seen near $\left\langle \hat{Q}_{40}\right\rangle =130\,\textrm{b}^{2}$,
and the fusion valley is near $\left\langle \hat{Q}_{40}\right\rangle =90\,\textrm{b}^{2}$.
For the $\left\langle Q_{20}\right\rangle =370\,\textrm{b}$ case,
only the fusion valley is observed, near $\left\langle \hat{Q}_{40}\right\rangle =140\,\textrm{b}^{2}$.}
\end{figure}

In Fig. \ref{cap:plot-300-and-370-ehfb}, we show a calculation of
the HFB energy for $^{240}\textrm{Pu}$ as a function of $Q_{40}$
($Q_{40}\equiv\left\langle \hat{Q}_{40}\right\rangle $) at two quadrupole
deformations, 300 b and 370 b, which correspond to the so-called cold
and hot fission limits, respectively \cite{berger84}. These calculations
were performed with 5 constraints (for the values of $\left\langle \hat{N}_{p}\right\rangle =94$,
$\left\langle \hat{N}_{n}\right\rangle =146$, $\left\langle \hat{Q}_{10}\right\rangle =0$,
$\left\langle \hat{Q}_{20}\right\rangle =300\,\textrm{b }$or $370\,\textrm{b }$,
and $80\,\textrm{b}^{2}\leq\left\langle \hat{Q}_{40}\right\rangle \leq200\,\textrm{b}^{2}$).
In the cold-fission case, a barrier of height $\sim4.0\,\textrm{MeV}$
relative to the fission-valley minimum separates the two valleys.
Near the hot-fission limit, the fission valley has disappeared and
the nucleus spontaneously falls into the fusion valley near $\left\langle \hat{Q}_{40}\right\rangle =140\,\textrm{b}^{2}$.
Between the hot and cold extremes, the nucleus can undergo fission
through a range of intermediate modes.

\begin{figure}
\includegraphics[%
  scale=0.33]{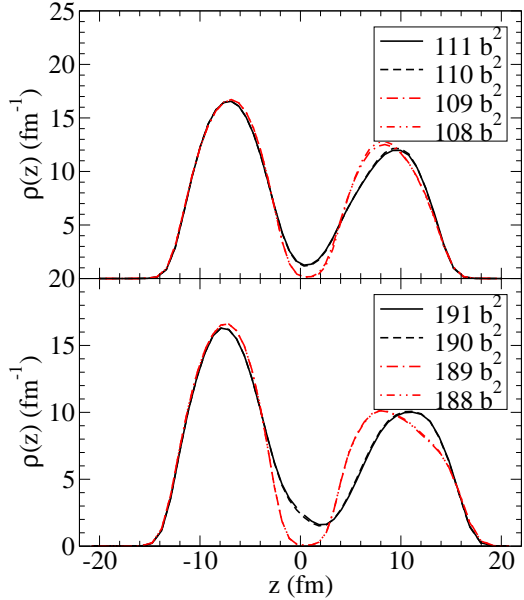}

\caption{\label{cap:300-and-370-breaking}(Color online) Calculated nuclear
densities in steps of $\Delta\left\langle \hat{Q}_{40}\right\rangle =1\,\textrm{b}^{2}$
around the scission configuration for cold (top panel) and hot (bottom
panel) fission. The legends give the values of $\left\langle \hat{Q}_{40}\right\rangle $
for the different curves.}
\end{figure}

The energy curves plotted in Fig. \ref{cap:plot-300-and-370-ehfb}
effectively represent slices at fixed values of $\left\langle \hat{Q}_{40}\right\rangle $
in Fig. 3 of \cite{berger84}. The most striking feature in Fig. \ref{cap:plot-300-and-370-ehfb}
is the sudden variation in energy over a very small step size in $\left\langle \hat{Q}_{40}\right\rangle $
of $1\,\textrm{b}^{2}$. In the cold-fission case, a drop of 2.7 MeV
is observed in going from $\left\langle \hat{Q}_{40}\right\rangle =110\,\textrm{b}^{2}$
to $109\,\textrm{b}^{2}$, and in the hot-fission case a more pronounced
drop of 7.6 MeV occurs in going from $\left\langle \hat{Q}_{40}\right\rangle =190\,\textrm{b}^{2}$
to $189\,\textrm{b}^{2}$. These abrupt changes in energy, which are
in contrast to the smooth behavior displayed in \cite{berger84},
correspond to a sudden reduction in the neck size (Fig. \ref{cap:300-and-370-breaking}),
which we take as an indicator of a transitional phase where the nucleus
is undergoing scission. Note that the identification of such transitional
phases requires extremely small variations of the constraints, which
could explain why they were not seen in \cite{berger84}. The precise
control of the constraints needed to study the region around scission
is one of the important points that emerges from the work presented
in this paper, and the motivation for going into some detail in the
description of the constraint-adjustment algorithm in the next section
and in appendix \ref{sec:Multiple-constraint-formalism}.

\begin{figure}
\includegraphics[%
  scale=0.33,
  angle=-90]{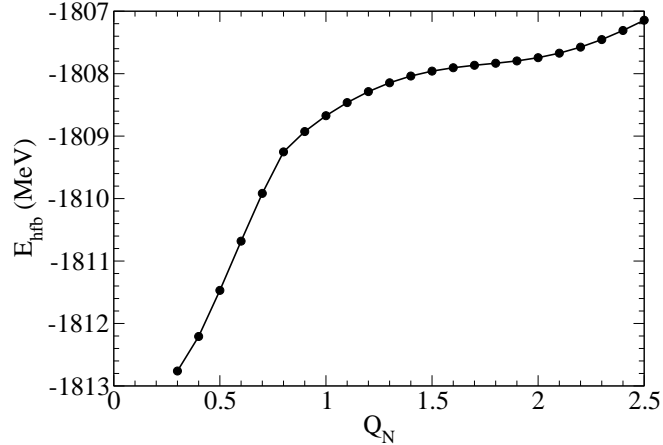}

\caption{\label{cap:plot-370-189-qn-ehfb}Variation of the HFB energy as a
function of the number of particles in the neck, defined by Eq. (\ref{eq:Qn-def}),
at the scission configuration ($\left\langle \hat{Q}_{40}\right\rangle =189\,\textrm{b}^{2}$)
for the hot-fission calculation ($\left\langle \hat{Q}_{20}\right\rangle =370\,\textrm{b}$)
in Fig. \ref{cap:plot-300-and-370-ehfb}.}
\end{figure}

The rapid change of the neck size mentioned above suggests the introduction
of a constraint proportional to the average number of particles $\left\langle \hat{Q}_{N}\right\rangle $
in the neck separating the nascent fragments, where \cite{warda02}\begin{eqnarray}
\hat{Q}_{N} & \equiv & \exp\left[-\frac{\left(z-z_{N}\right)^{2}}{a_{N}^{2}}\right]\label{eq:Qn-def}\end{eqnarray}
with $a_{N}=1\,\textrm{fm}$, and $z_{N}$ is the position of the
neck (defined as the point between the fragments where the matter
density is lowest). As shown in Fig. \ref{cap:plot-370-189-qn-ehfb},
the energy calculated as a function of $\left\langle \hat{Q}_{N}\right\rangle $
becomes smoother and continuous. A more detailed discussion of this
result is given in the latter part of section \ref{sub:The-HFB-convergence}.

\subsection{The HFB convergence algorithm\label{sub:The-HFB-convergence}}

The control of HFB calculations with multiple constraints is a delicate
procedure, made difficult by the number of constraints and their inherent
correlations. Because the topic continues to be of current interest
in problems that rely on constrained-HFB methods even beyond fission
\cite{baran08,egido95}, the convergence algorithm used in the present
HFB calculations is discussed in detail here. The algorithm must balance,
at each iteration, the diagonalization of the HFB Hamiltonian to ensure
self-consistency, and adjustment of the Lagrange multipliers in Eq.
(\ref{eq:constrd-ham}). The main steps of the algorithm are as follows

\begin{enumerate}
\item Read initial generalized density $R$ and Lagrange multipliers $\lambda_{i}$
\item \label{enu:Construct-constrained-HFB}Construct constrained HFB Hamiltonian
$\mathcal{H}\left(R,\left\{ \lambda_{i}\right\} \right)$
\item Diagonalize $\mathcal{H}\left(R,\left\{ \lambda_{i}\right\} \right)$\label{enu:Diagonalize-H}
\item Construct new $R$\label{enu:Construct-new-R}
\item \label{enu:Mix-R-between}Mix $R$ between consecutive iterations
using a mixing parameter $\alpha$ (see Eq. (\ref{eq:mixing-prescription}))
\item \label{enu:Adjust-value-of}Adjust value of $\alpha$ based on convergence
criterion
\item \label{enu:Calculate-deltalambdai-needed}Calculate $\delta\lambda_{i}$
needed to yield desired constraint values, adjust $\lambda_{i}$
\item \label{enu:Calculate-deltaR}Calculate $\delta R$ corresponding to
the $\delta\lambda_{i}$, adjust $R$
\item \label{enu:If-HFB-solution}If HFB solution is not converged, return
to step \ref{enu:Construct-constrained-HFB}
\end{enumerate}
The first 4 steps in this algorithm are fairly self-explanatory and
make use of the formalism derived in section \ref{sub:General-HFB-formalism}.
We will examine the remaining steps in greater detail since they are
not typically discussed in depth in the literature. 

At the end of each iteration $i$, the convergence of the HFB solution
is assessed by calculating the largest variation from the previous
iteration in the elements of the generalized density matrix,\begin{eqnarray}
\varepsilon_{i} & \equiv & \sup\left|R_{mn}^{pq}\left(i\right)-R_{mn}^{pq}\left(i-1\right)\right|\label{eq:conv-eps-def}\end{eqnarray}
The quantity $\varepsilon_{i}$ is also used to determine the coefficient
$\alpha$ in step \ref{enu:Mix-R-between} which mixes the generalized
densities between successive iterations using an adjustable coefficient
$\alpha$,\begin{eqnarray}
R_{mn}^{pq}\left(i\right) & \rightarrow & \left(1-\alpha\right)R_{mn}^{pq}\left(i\right)+\alpha R_{mn}^{pq}\left(i-1\right)\label{eq:mixing-prescription}\end{eqnarray}
with $0\leq\alpha\leq1$. This mixing is essential to slow down the
convergence algorithm which would otherwise often behave erratically
in the first few iterations and could fail to converge at all. The
mixing coefficient $\alpha$ is adjusted in step \ref{enu:Adjust-value-of}
in such a way that it tends to zero as $\varepsilon_{i}$ decreases.
In practice, two thresholds are supplied, $\varepsilon_{min}$ and
$\varepsilon_{max}$, along with a maximum value $\alpha_{max}$ for
the mixing coefficient such that\begin{eqnarray*}
\alpha & = & \begin{cases}
\alpha_{max} & \varepsilon_{i}\geq\varepsilon_{max}\\
\alpha_{max}\frac{\varepsilon_{i}-\varepsilon_{min}}{\varepsilon_{max}-\varepsilon_{min}} & \varepsilon_{min}<\varepsilon_{i}<\varepsilon_{max}\\
0 & \varepsilon_{i}\leq\varepsilon_{min}\end{cases}\end{eqnarray*}
Furthermore, if the HFB solution diverges from one iteration to the
next (i.e., if $\varepsilon_{i}>\varepsilon_{i-1}$) then $\alpha$
is set to $\alpha_{max}$ and remains at that value until the HFB
solution converges again. For the work in this paper we have used
$\varepsilon_{min}=10^{-3}$ or $10^{-4}$, $\varepsilon_{max}=10^{-1}$,
and $\alpha_{max}=0.5$ (or in a few cases $0.8$ for a slower initial
convergence). We note in passing that the mixing of generalized density
matrices is a global operation, i.e. the same coefficient $\alpha$
is used for all the matrix elements. The Broyden method, or its more
elaborate modified version \cite{baran08}, could provide a better
alternative for optimizing the choice of the mixing coefficient by
associating an independent value of $\alpha$ to each matrix element.

The formalism needed to adjust the Lagrange parameters in step \ref{enu:Calculate-deltalambdai-needed},
and the generalized density in step \ref{enu:Calculate-deltaR} is
presented in appendix \ref{sec:Multiple-constraint-formalism}, and
we stress the importance of adjusting both for a stable convergence
of the HFB method. The algorithm is considered to have converged in
step \ref{enu:If-HFB-solution} if $\varepsilon_{i}\leq\varepsilon_{min}$
for several iterations (typically 2 in the present work).

\begin{figure}
\includegraphics[%
  scale=0.33,
  angle=-90]{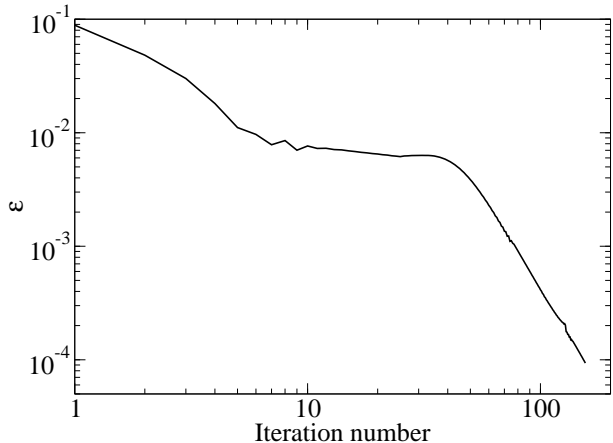}

\caption{\label{cap:conv300-110-supd}Plot of the convergence metric, given
by Eq. (\ref{eq:conv-eps-def}), as a function of HFB iteration number
for the $\left\langle \hat{Q}_{40}\right\rangle =110\,\textrm{b}^{2}$
cold-fission point in Fig. \ref{cap:plot-300-and-370-ehfb}.}
\end{figure}

In order to illustrate various aspects of the convergence algorithm,
we have examined the cold-fission point at $\left\langle \hat{Q}_{40}\right\rangle =110\,\textrm{b}^{2}$
in Fig. \ref{cap:plot-300-and-370-ehfb} in detail. Because this point
corresponds to a local maximum in the HFB energy, its calculation
is particularly demanding on the convergence algorithm. In Fig. \ref{cap:conv300-110-supd}
we show the convergence criterion, $\varepsilon$, calculated using
Eq. (\ref{eq:conv-eps-def}) at each iteration. The HFB solution is
found to better than $\varepsilon<10^{-4}$ after 156 iterations in
this case. We note a region in Fig. \ref{cap:conv300-110-supd} roughly
between iterations 10 and 40, where $\varepsilon$ appears to be relatively
constant and the convergence is correspondingly slow. In this region,
all the constraints appear to be close to their desired values, except
for the dipole moment. The $\left\langle \hat{Q}_{10}\right\rangle $
value is still relatively large ($\sim0.06-0.2$ fm) and may be responsible
for the stagnant convergence.

\begin{figure}
\includegraphics[%
  scale=0.33,
  angle=-90]{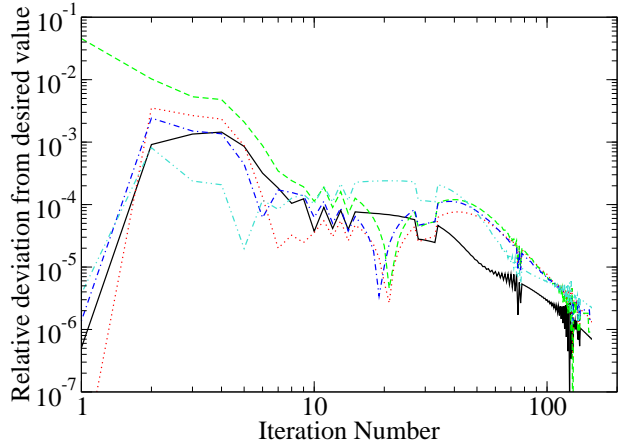}

\caption{\label{cap:conv300-110-reldev-fig}(Color online) Relative deviations
of the calculated constraint values from their desired values as a
function of HFB iteration number for the calculation with $\left\langle \hat{Q}_{40}\right\rangle =110\,\textrm{b}^{2}$.
The relative deviation for the dipole moment is given by Eq. (\ref{eq:reldev-Q10}),
and by Eq. (\ref{eq:reldev-Q}) for all other constraints. The constraints
shown are: $\left\langle \hat{Q}_{10}\right\rangle $ (black solid
line), $\left\langle \hat{Q}_{20}\right\rangle $ (red dotted line),
$\left\langle \hat{Q}_{40}\right\rangle $ (green dashed line), $\left\langle \hat{N}_{n}\right\rangle $
(blue dot-dashed line), and $\left\langle \hat{N}_{p}\right\rangle $
(turquoise dot-dot-dashed line).}
\end{figure}

In Fig. \ref{cap:conv300-110-reldev-fig} we examine the adjustment
of the five constraints at each iteration. The figure shows the relative
deviation of each constraint from the desired value. For all but the
dipole-moment constraint, this relative deviation of the calculated
average value $\left\langle \hat{Q}\right\rangle $ of the constraint
from its desired value $q$ is given by\begin{equation}
\left|\frac{\left\langle \hat{Q}\right\rangle -q}{q}\right|\label{eq:reldev-Q}\end{equation}
In the case of the dipole moment, the desired value is $q_{10}=0$
and Eq. (\ref{eq:reldev-Q}) cannot be used. Instead, we obtain from
$\left\langle \hat{Q}_{10}\right\rangle $ the position of the centroid
of the nucleus, given by $\left\langle \hat{Q}_{10}\right\rangle /A$
where $A=240$ is the total number of nucleons, and compare it to
the calculated root-mean-squared radius of the nucleus, $R_{rms}$,
using the ratio\begin{equation}
\left|\frac{\left\langle \hat{Q}_{10}\right\rangle }{AR_{rms}}\right|\label{eq:reldev-Q10}\end{equation}

The calculation is started from an HFB solution that differs only
in the value of the hexadecapole constraint, $\left\langle \hat{Q}_{40}\right\rangle =115\,\textrm{b}^{2}$,
with all other constraints the same. Hence we see in Fig. \ref{cap:conv300-110-reldev-fig}
that at the first iteration, all relative deviations except the one
for the hexadecapole-moment constraint are small. The calculation
converges to the desired level of accuracy after 156 iterations.

\begin{figure}
\includegraphics[%
  scale=0.33,
  angle=-90]{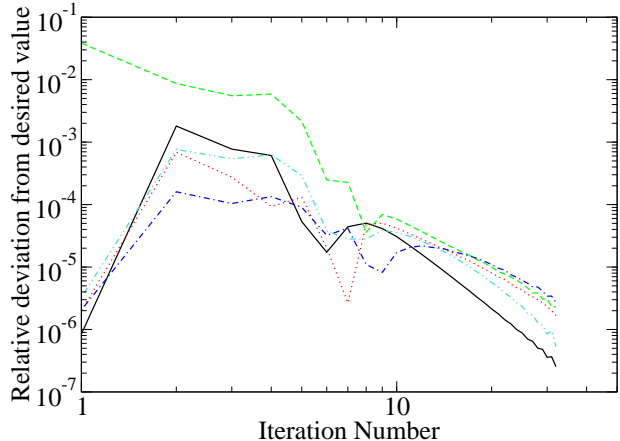}

\caption{\label{cap:conv300-130-reldev-fig}(Color online) Same as Fig. \ref{cap:conv300-110-reldev-fig},
but for the calculation with $\left\langle \hat{Q}_{40}\right\rangle =130\,\textrm{b}^{2}$.}
\end{figure}

This difficult convergence should be contrasted with the calculation
of the cold-fission point at $\left\langle \hat{Q}_{40}\right\rangle =130\,\textrm{b}^{2}$,
near the bottom of the fission valley in Fig. \ref{cap:plot-300-and-370-ehfb}.
The relative deviations of the constraints for this more stable calculation
are shown in Fig. \ref{cap:conv300-130-reldev-fig}. After the tenth
iteration, all constraints tend to the desired value rapidly and smoothly.
This calculation is converged to the same level of accuracy as the
one at $\left\langle \hat{Q}_{40}\right\rangle =110\,\textrm{b}^{2}$
after only 33 iterations.

Finally, we discuss in greater detail the discontinuities observed
in Fig. \ref{cap:plot-300-and-370-ehfb}. Such discontinuities have
been alluded to in the literature \cite{moller04} as a potential
difficulty for microscopic calculations. In this section, we show
how these discontinuities are an indicator of a change in the meaning
of certain collective coordinates near the critical scission configurations.
We also show how these discontinuities can be eliminated through the
choice of a more appropriate collective coordinate.

\begin{figure}
\includegraphics[%
  scale=0.33,
  angle=-90]{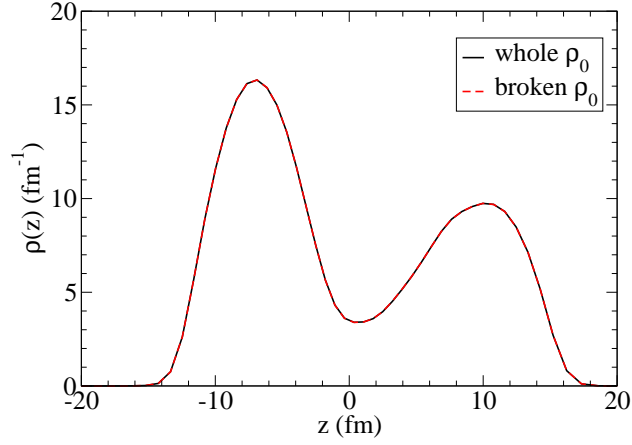}

\caption{\label{cap:300-130-broken}(Color online) Comparison of nuclear densities
for the $\left\langle \hat{Q}_{40}\right\rangle =130\,\textrm{b}^{2}$
cold-fission point in Fig. \ref{cap:plot-300-and-370-ehfb}, starting
either from a whole (solid black line) or scissioned/broken (dashed
red line) initial configuration of the nuclear density in the HFB
iterations. All moments up to the hexadecapole have been constrained
to the same values for the two calculations.}
\end{figure}

\begin{figure}
\includegraphics[%
  scale=0.33,
  angle=-90]{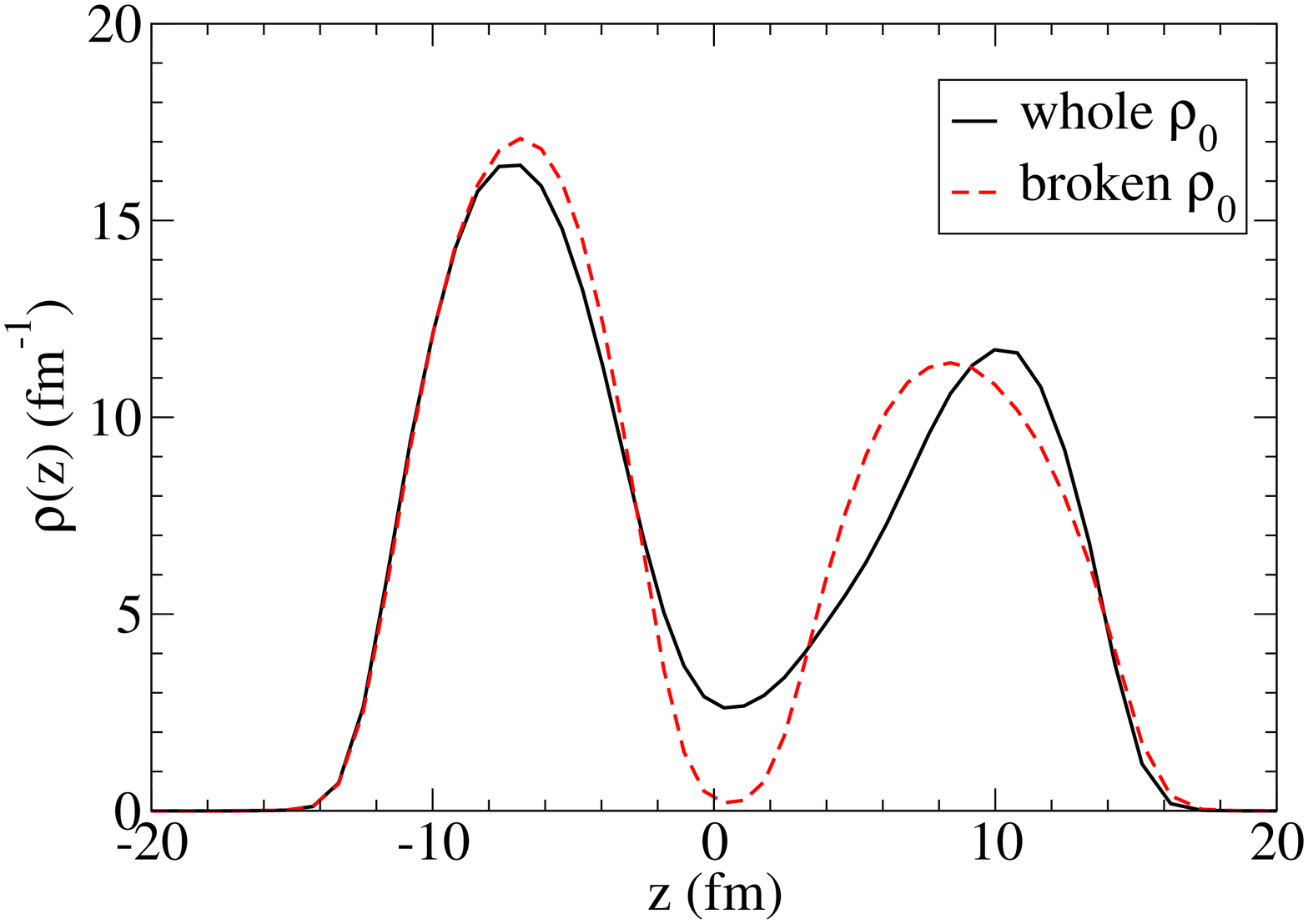}

\caption{\label{cap:300-115-broken}(Color online) Same as Fig. \ref{cap:300-130-broken},
but for a calculation at $\left\langle \hat{Q}_{40}\right\rangle =115\,\textrm{b}^{2}$.}
\end{figure}

\begin{figure}
\includegraphics[%
  scale=0.33,
  angle=-90]{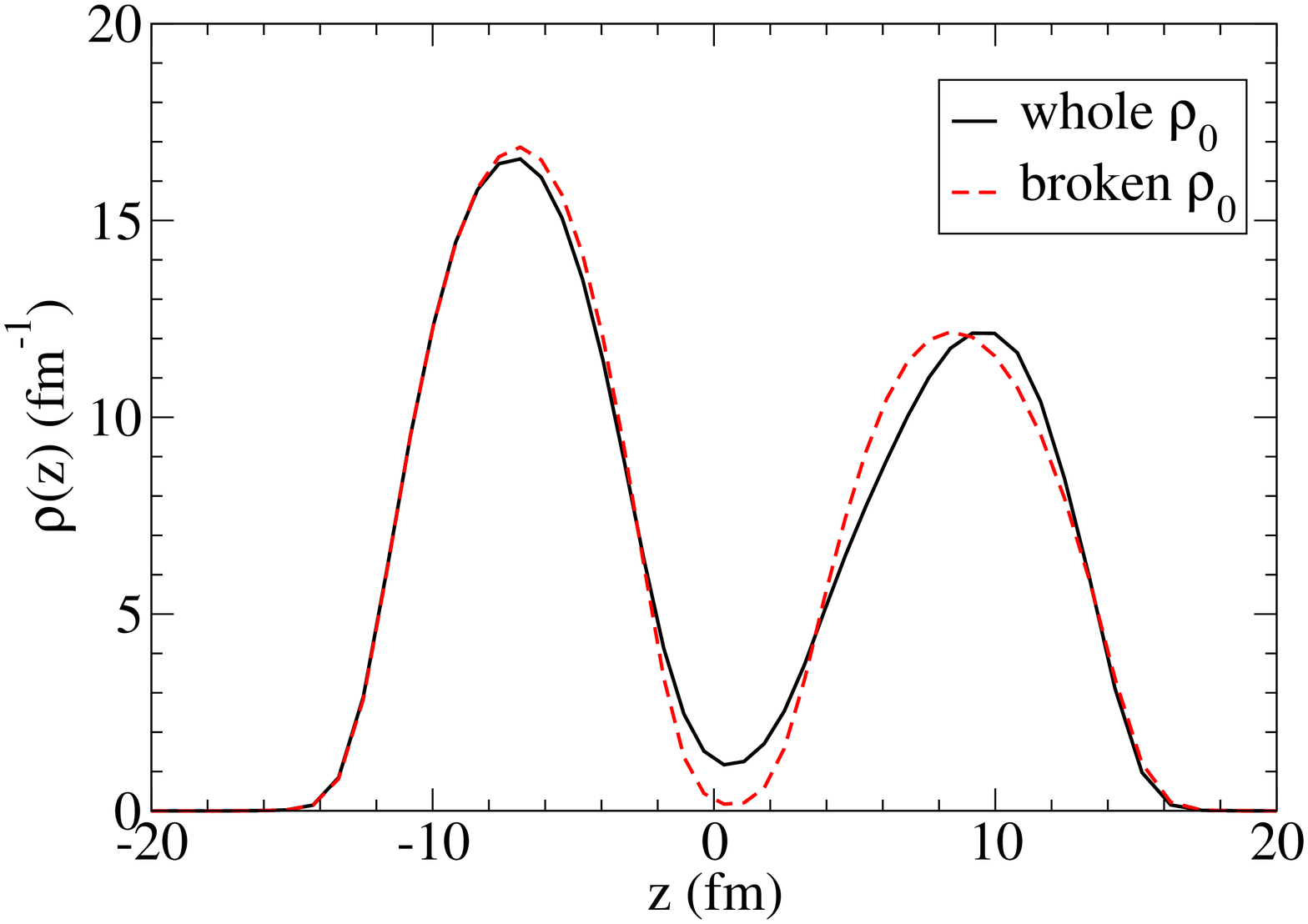}

\caption{\label{cap:300-110-broken}(Color online) Same as Fig. \ref{cap:300-130-broken},
but for a calculation at $\left\langle \hat{Q}_{40}\right\rangle =110\,\textrm{b}^{2}$.}
\end{figure}

The impact of these discontinuities can be felt even before the scission
configuration is reached. We illustrate this point by showing the
results of HFB calculations, performed with identical multipole constraints
up to the hexadecapole moment (i.e., with the same $\left\langle \hat{Q}_{10}\right\rangle $,
$\left\langle \hat{Q}_{20}\right\rangle $, $\left\langle \hat{Q}_{30}\right\rangle $,
$\left\langle \hat{Q}_{40}\right\rangle $ values), but different
initial densities. We will approach the cold-fission scission configuration
near $\left\langle \hat{Q}_{40}\right\rangle =110\,\textrm{b}^{2}$
in Fig. \ref{cap:plot-300-and-370-ehfb} with an initial density corresponding
to either a scissioned or non-scissioned nucleus. The first calculation,
shown in Fig. \ref{cap:300-130-broken}, was performed at $\left\langle \hat{Q}_{40}\right\rangle =130\,\textrm{b}^{2}$,
near the bottom of the fission valley. Two curves are shown, corresponding
to a initial choice of the generalized density calculated at $\left\langle \hat{Q}_{40}\right\rangle =135\,\textrm{b}^{2}$
(whole nucleus), and $\left\langle \hat{Q}_{40}\right\rangle =90\,\textrm{b}^{2}$
(broken/scissioned nucleus). As expected, both choices of starting
point lead to exactly the same HFB solution, as is evidenced by the
overlapping density curves in Fig. \ref{cap:300-130-broken}. By contrast,
Fig. \ref{cap:300-115-broken} compares calculations at $\left\langle \hat{Q}_{40}\right\rangle =115\,\textrm{b}^{2}$
(i.e., near scission), starting from solutions at $\left\langle \hat{Q}_{40}\right\rangle =120\,\textrm{b}^{2}$
(whole) and $\left\langle \hat{Q}_{40}\right\rangle =90\,\textrm{b}^{2}$
(broken). Both solutions have the same values of the first four moments,
yet the calculation started from a whole solution leads to a whole
result, while the broken starting configuration leads to a broken-nucleus
solution. A similar effect is observed in Fig. \ref{cap:300-110-broken},
corresponding to a calculation very close to scission at $\left\langle \hat{Q}_{40}\right\rangle =110\,\textrm{b}^{2}$
with starting densities from $\left\langle \hat{Q}_{40}\right\rangle =115\,\textrm{b}^{2}$
(whole) and $\left\langle \hat{Q}_{40}\right\rangle =90\,\textrm{b}^{2}$
(broken) solution. Note that these HFB calculations are performed
with an unprecedented 7 simultaneous constraints.

\begin{figure}
\includegraphics[%
  scale=0.35]{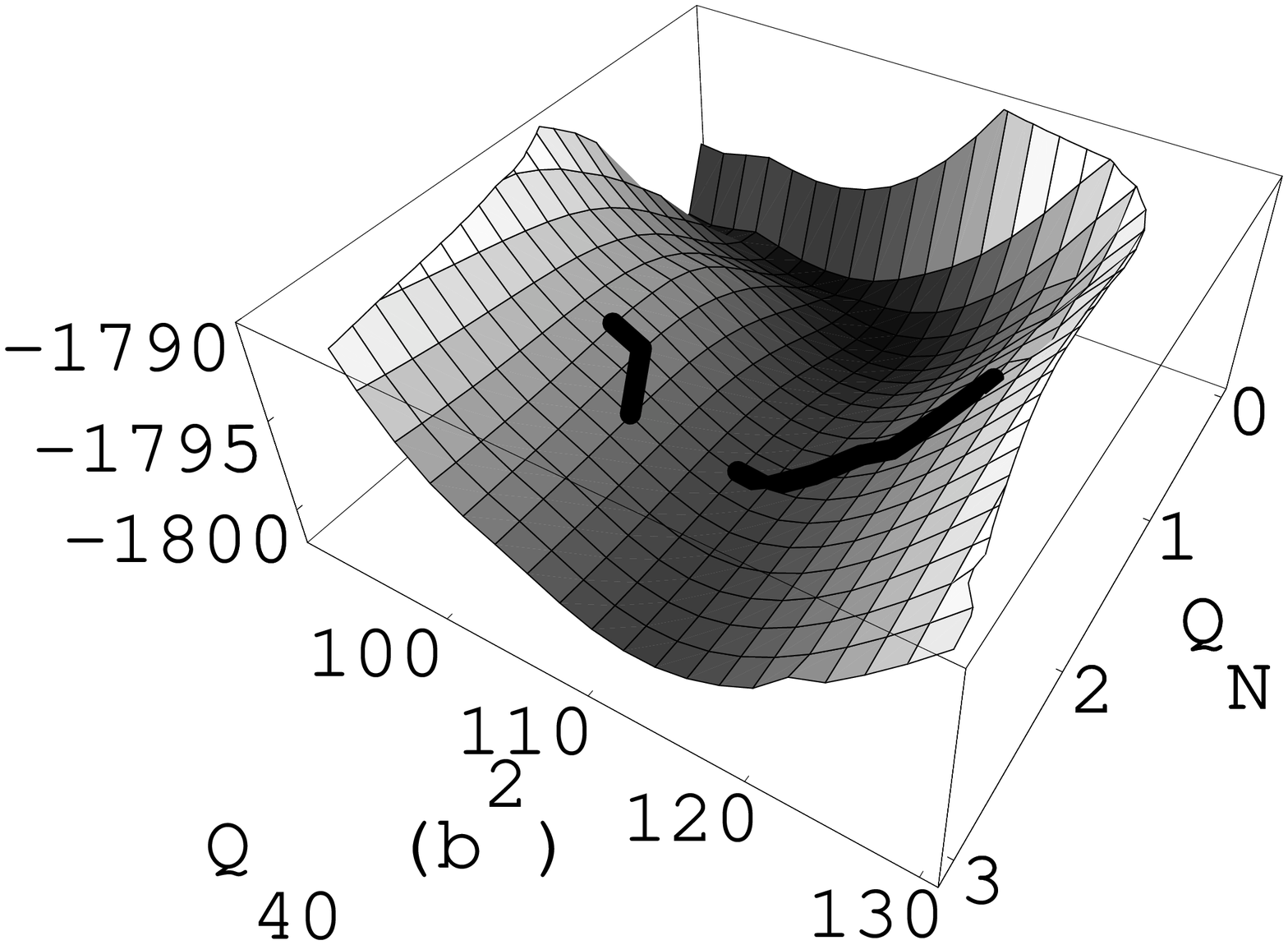}

\caption{\label{cap:map4n-300-034}Energy surface calculated with constraints
on $\left\langle \hat{N}_{n}\right\rangle =146$, $\left\langle \hat{N}_{p}\right\rangle =94$,
$\left\langle \hat{Q}_{10}\right\rangle =0$, $\left\langle \hat{Q}_{20}\right\rangle =300\,\textrm{b}$,
$\left\langle \hat{Q}_{30}\right\rangle =34.951\,\textrm{b}^{3/2}$,
$90\,\textrm{b}^{3/2}\leq\left\langle \hat{Q}_{40}\right\rangle \leq130\,\textrm{b}^{3/2}$,
and $0.05\leq\left\langle \hat{Q}_{N}\right\rangle \leq3.05$. The
dark lines along the surface mark the position of a small local barrier
on the surface.}
\end{figure}

The densities plotted in Figs. \ref{cap:300-130-broken}-\ref{cap:300-110-broken}
reveal a complex relationship between the hexadecapole and $Q_{N}$
degrees of freedom. These two coordinates are not related by a one-to-one
mapping and cannot be used interchangeably to drive the system to
scission. In Fig. \ref{cap:map4n-300-034} we show the HFB energy
surface as a function of $Q_{40}$ and $Q_{N}$ for the calculation
with all moments up to hexadecapole fixed. In particular, $\left\langle \hat{Q}_{20}\right\rangle =300\,\textrm{b}$,
and $\left\langle \hat{Q}_{30}\right\rangle =34.951\,\textrm{b}^{3/2}$--the
value of the octupole moment for the two calculations in Fig. \ref{cap:300-110-broken}.
The shape of the energy surface suggests that energy-minimizing HFB
solutions can exist which have the same value of $\left\langle \hat{Q}_{40}\right\rangle $,
but distinct values of $\left\langle \hat{Q}_{N}\right\rangle $.
For most--but not all--values of $\left\langle \hat{Q}_{40}\right\rangle $
a small barrier in the surface (marked by a solid line along the surface
in the figure) separates the minima with differing values of $\left\langle \hat{Q}_{N}\right\rangle $.
This barrier is at best a few hundred keV's high and decreases rapidly
with decreasing $\left\langle \hat{Q}_{40}\right\rangle $ as we approach
the scission configuration. At $\left\langle \hat{Q}_{40}\right\rangle =110\,\textrm{b}^{2}$
the barrier has dropped to only 1.8 keV and vanishes completely between
$\left\langle \hat{Q}_{40}\right\rangle =104\,\textrm{b}^{2}$ and
$110\,\textrm{b}^{2}$. This break in the barrier causes the discontinuity
in Fig. \ref{cap:plot-300-and-370-ehfb}, where the calculations are
performed without a constraint on $\left\langle \hat{Q}_{N}\right\rangle $
to prevent the HFB calculation from falling into the scissioned configuration.

Near scission, the total multipole moments of the nucleus are determined
by the intrinsic and relative moments of the fragments, and rearrangements
between these terms can produce different matter distributions with
the same overall moments, at least up to the hexadecapole. Thus imposing
a constraint on $\left\langle \hat{Q}_{40}\right\rangle $ will not
necessarily result in a constraint on the neck size near scission.
The $\left\langle \hat{Q}_{N}\right\rangle $ constraint on the other
hand was already shown to produce a smooth energy dependence in Fig.
\ref{cap:plot-370-189-qn-ehfb} and is therefore the suitable coordinate
in the study of fission for configurations near and beyond scission.

\subsection{Scission in the constrained-HFB approach}

In this section, we briefly discuss various signatures of scission.
Some of the characteristics of scission have already been mentioned
in sections \ref{sub:Multiple-constraints-in} and \ref{sub:The-HFB-convergence}.
The standard indicators of scission are sudden changes in either energy
(interaction energy between fragments or total HFB energy) or shape
(neck size or hexadecapole moment) for the nucleus \cite{dubray08}.
For the work in this paper, we use the same semiclassical definition
of the nascent fission fragments as in \cite{dubray08}, where a position
along the symmetry axis of the nucleus is identified as a divider
between left and right fragments, and the fragment properties are
obtained as integrals over the density with this cut as an endpoint
for the integrals. In a forthcoming publication \cite{younes09a},
we will adopt a more microscopic criterion to identify the fragment
\cite{younes09b}, based on the individual single-particle wave functions,
and using the changes in the interaction energy between fragments
as an indicator of scission. In this paper we will focus instead on
the HFB energy and the number of particles in the neck before and
after scission.

Consider, for example, the cold-fission calculation in Fig. \ref{cap:plot-300-and-370-ehfb}.
At $\left\langle \hat{Q}_{40}\right\rangle =110\,\textrm{b}^{2}$
there is still a significant amount of matter in the neck connecting
the nascent fragment with $\left\langle \hat{Q}_{N}\right\rangle =2.41$.
At $\left\langle \hat{Q}_{40}\right\rangle =109\,\textrm{b}^{2}$
however, the neck breaks and $\left\langle \hat{Q}_{N}\right\rangle $
drops to $0.50$ particles. This sudden variation in shape over a
small increment in hexadecapole moment is shown in the top panel of
Fig. \ref{cap:300-and-370-breaking}. At $\left\langle \hat{Q}_{40}\right\rangle =90\,\textrm{b}^{2}$,
the bottom of the fusion valley, $\left\langle \hat{Q}_{N}\right\rangle $
has been reduced to $0.09$ particles. From $\left\langle \hat{Q}_{40}\right\rangle =110\,\textrm{b}^{2}$
to $109\,\textrm{b}^{2}$, the total HFB energy drops by 2.7 MeV,
and the difference in energy between $\left\langle \hat{Q}_{40}\right\rangle =110\,\textrm{b}^{2}$
and $90\,\textrm{b}^{2}$ is 10.2 MeV.

A similar analysis can be performed for the hot-fission calculation
in Fig. \ref{cap:plot-300-and-370-ehfb}. In this case, the last point
where a sizable neck still exists between the nascent fragment is
at $\left\langle \hat{Q}_{40}\right\rangle =190\,\textrm{b}^{2}$,
with $\left\langle \hat{Q}_{N}\right\rangle =2.92$ particles. By
$\left\langle \hat{Q}_{40}\right\rangle =189\,\textrm{b}^{2}$ the
neck has essentially disappeared, and $\left\langle \hat{Q}_{N}\right\rangle $
has dropped to 0.23 particles. At the bottom of the fusion valley,
where $\left\langle \hat{Q}_{40}\right\rangle =140\,\textrm{b}^{2}$,
there are only $\left\langle \hat{Q}_{N}\right\rangle =0.02$ particles
in the neck. The change in shape is plotted in the bottom panel of
Fig. \ref{cap:300-and-370-breaking}. The drops in energy are more
significant than in the cold-fission case. From $\left\langle \hat{Q}_{40}\right\rangle =190\,\textrm{b}^{2}$
to $189\,\textrm{b}^{2}$, the total HFB energy drops by 7.6 MeV,
and from $\left\langle \hat{Q}_{40}\right\rangle =190\,\textrm{b}^{2}$
to $140\,\textrm{b}^{2}$, it drops by 20.1 MeV.

\section{Results\label{sec:Results}}

\subsection{Benchmark: $^{226}\textrm{Th}$ scission\label{sub:Benchmark:226Th}}

We have performed HFB calculations of hot-fission properties for $^{226}\textrm{Th}$,
in order to compare with the results in \cite{dubray08} that were
obtained with two-center HFB calculations. We have used both the basis
truncation of Eq. (\ref{eq:egido-trunc}) with $N=13$ and $q=1.5$,
and the one given by Eq. (\ref{eq:flocard-trunc}) with $N=13$. The
oscillator-frequency parametrization of Eqs. (\ref{eq:parm-hw0})
and (\ref{eq:parm-q}) was used, even though it was obtained for calculations
in $^{240}\textrm{Pu}$. We will show that our results are in good
agreement with those of Dubray \emph{et al}. \cite{dubray08} for
$^{226}\textrm{Th}$ with either basis truncation scheme.

\begin{figure}
\includegraphics[%
  scale=0.33,
  angle=-90]{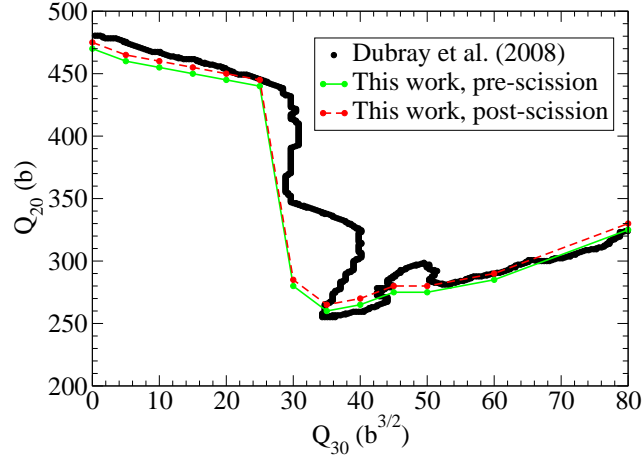}

\caption{\label{cap:226th-d1s-scl23}(Color online) Scission line for $^{226}\textrm{Th}$
obtained in this work, and compared to the result of Dubray \emph{et
al}. \cite{dubray08}. The solid disks connected by a solid green
line represent HFB solutions just before scission in this work, and
the solid disks connected by a dashed red line represent solutions
immediately after scission in this work. The thick solid black curve
is the scission line taken from \cite{dubray08}.}
\end{figure}

\begin{figure}
\includegraphics[%
  scale=0.33,
  angle=-90]{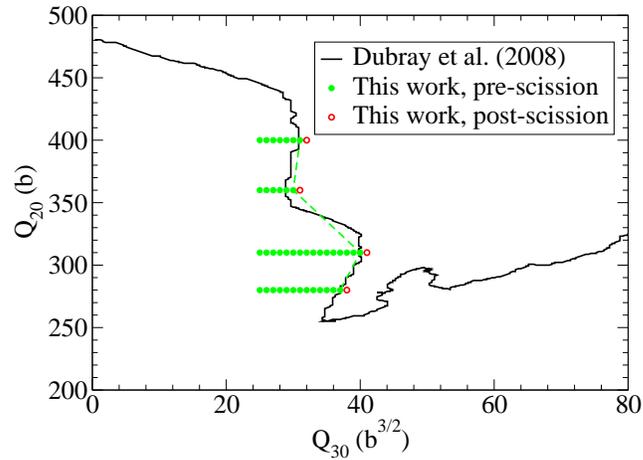}

\caption{\label{cap:226th-scl23-horiz}(Color online) Large-basis HFB calculations
in $^{226}\textrm{Th}$ along lines with fixed $\left\langle \hat{Q}_{20}\right\rangle $
performed to reproduce the details of the scission line found in Dubray
\emph{et al}. \cite{dubray08}. A dashed line connects the last point
before scission, and should be compared to the Dubray \emph{et al}.
result (solid line).}
\end{figure}

In Fig. \ref{cap:226th-d1s-scl23}, we plot the hot-scission line
for $^{226}\textrm{Th}$, and compare it to the one obtained in \cite{dubray08}.
The scission line was determined by performing series of calculations
at fixed $\left\langle \hat{Q}_{30}\right\rangle $ and increasing
values of $\left\langle \hat{Q}_{20}\right\rangle $ by 5 b, each
calculation using the previous one as a starting point, until an HFB
solution was found where the neck size decreased drastically. Lines
separated by $\Delta\left\langle \hat{Q}_{20}\right\rangle $= 5 b
connecting the HFB solutions just before and just after the breaking
of the neck are displayed in Fig. \ref{cap:226th-d1s-scl23}, bracketing
the actual scission line. These lines are in good agreement with the
$^{226}\textrm{Th}$ scission line in \cite{dubray08}. In Fig. \ref{cap:226th-scl23-horiz}
we examine the region with $\left\langle \hat{Q}_{30}\right\rangle =25-35\;\textrm{b}^{3/2}$
in greater detail. A series of HFB calculations were performed at
constant $\left\langle \hat{Q}_{20}\right\rangle $ values of 280,
310, 360, and 400 b starting from $\left\langle \hat{Q}_{30}\right\rangle =25\;\textrm{b}^{3/2}$
in each case and proceeding in steps of $\Delta\left\langle \hat{Q}_{30}\right\rangle =1\;\textrm{b}^{3/2}$.
For these calculations, the basis truncation of Eq. (\ref{eq:flocard-trunc})
was used with $N=13$ in order to provide a larger number of oscillator
shells (up to 26 in practice) in the $z$ direction, while keeping
the overall number of basis states relatively low. With these large-basis
calculations, we find that the results of Dubray \emph{et al}. \cite{dubray08}
are very well reproduced.

\begin{figure}
\includegraphics[%
  scale=0.33,
  angle=-90]{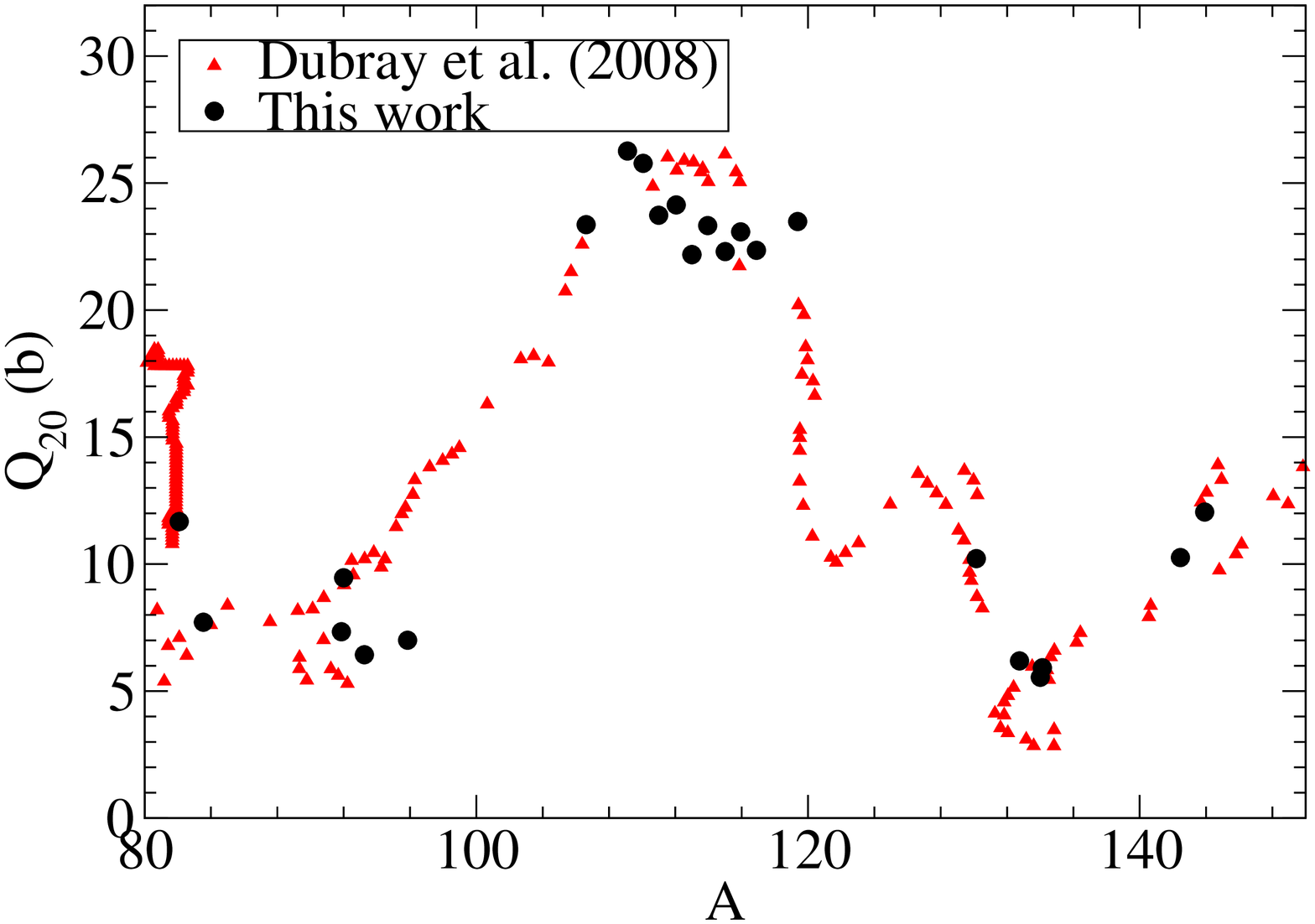}

\caption{\label{cap:226th-a-q20}(Color online) Comparison of fission-fragment
quadrupole moments as a function of fragment mass number between this
work (solid black disks) and the results in \cite{dubray08} (solid
red triangles).}
\end{figure}

\begin{figure}
\includegraphics[%
  scale=0.33,
  angle=-90]{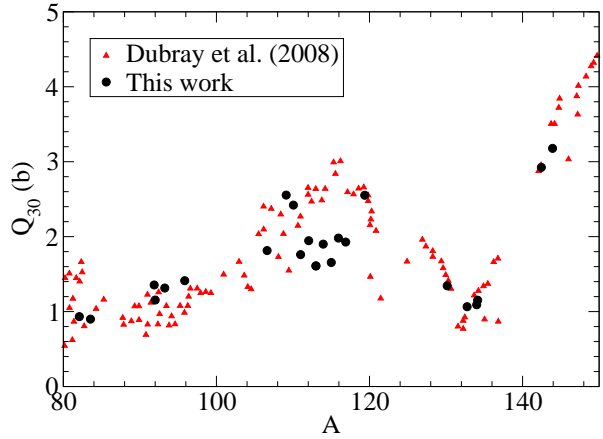}

\caption{\label{cap:226th-a-q30}(Color online) Same as Fig. \ref{cap:226th-a-q20},
but for the fission-fragment octupole moments.}
\end{figure}

In Fig. \ref{cap:226th-a-q20}, we compare the mass quadrupole moment
calculated for the fragments for the HFB solutions just before scission
(solid disks connected by solid lines in Fig. \ref{cap:226th-d1s-scl23})
to the corresponding result in \cite{dubray08}. As in \cite{dubray08},
the $Q_{20}$ values were calculated by integration over the left-
and right-fragment densities, truncated at the neck position. The
results of \cite{dubray08} are well reproduced by our calculations.
Similarly, in Fig. \ref{cap:226th-a-q30}, we show the octupole moment
of the fragments compared to the Dubray \emph{et al}. results. In
this case as well, the agreement between the two sets of calculations
is good.

The agreement between one-center and two-center calculations in Figs.
(\ref{cap:226th-d1s-scl23})-(\ref{cap:226th-a-q30}) is reassuring,
both as a benchmark for the HFB code used in this work, and as an
assessment of the applicability of the one-center basis near scission.
With these results in mind, we turn next to the fission properties
of $^{240}\textrm{Pu}$.

\subsection{$^{240}\textrm{Pu}$ scission}

For the $^{240}\textrm{Pu}$ calculations, we have used the truncation
scheme of Eq. (\ref{eq:flocard-trunc}) with $N=13$. The parameterization
in Eqs. (\ref{eq:parm-hw0}) and (\ref{eq:parm-q}) was adopted for
the HO frequencies.

\begin{figure}
\includegraphics[%
  scale=0.33]{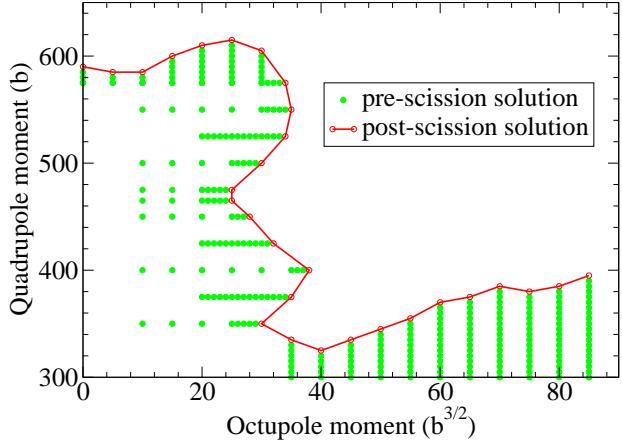}

\caption{\label{cap:240pu-scl23}(Color online) Scission line for $^{240}\textrm{Pu}$
obtained in this work. All calculations were done using the basis
truncation of Eq. (\ref{eq:flocard-trunc}). The solid green disks
represent HFB calculations producing a whole (non-scissioned) nuclear
density. The empty red circles connected by a solid line represent
scissioned configurations.}
\end{figure}

Fig. \ref{cap:240pu-scl23} illustrates the search for the hot-scission
line in $^{240}\textrm{Pu}$. Points along lines with fixed $\left\langle \hat{Q}_{30}\right\rangle $
or $\left\langle \hat{Q}_{20}\right\rangle $ increasing in steps
of $1\,\textrm{b}^{3/2}$ and 5 b near the scission line, respectively,
denote individual HFB calculations, each using the previous one as
a starting point. As in the case of $^{226}\textrm{Th}$ in Fig. \ref{cap:226th-d1s-scl23},
the nucleus tends to stretch to much larger deformations in the symmetric
limit. This leads to fragments that are formed much further apart
in symmetric fission, and a corresponding drop in their mutual Coulomb
repulsion--and therefore their total kinetic energy--as observed experimentally
\cite{wagemans84}. As in the case of $^{226}\textrm{Th}$, we also
observe regions around $Q_{20}=550\,\textrm{b}/Q_{30}=35\,\textrm{b}^{3/2}$
and $Q_{20}=400\,\textrm{b}/Q_{30}=38\,\textrm{b}^{3/2}$ where the
scission line {}``bulges out''. In these regions, for a given $Q_{30}$
value, the nucleus may scission at more than one value of $Q_{20}$.

\begin{figure}
\includegraphics[%
  scale=0.33,
  angle=-90]{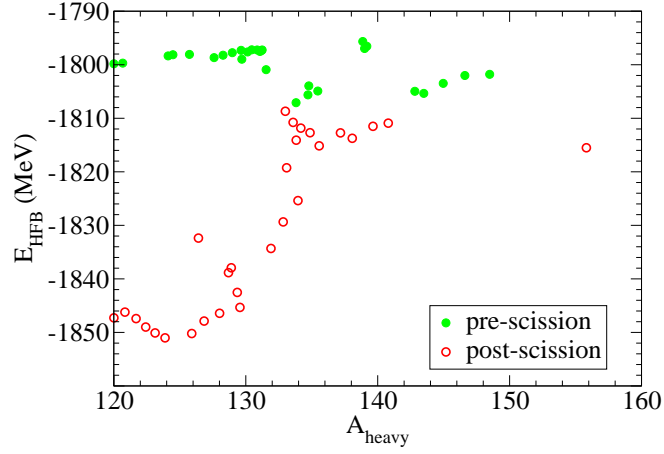}

\caption{\label{cap:240pu-frag-ehfb}(Color online) HFB energy of the fissioning
nucleus, plotted as a function of the heavy-fragment mass number,
obtained from the HFB calculations just before (solid green disks)
and just after (empty red circles) scission in Fig. \ref{cap:240pu-scl23}.}
\end{figure}

Fig. \ref{cap:240pu-frag-ehfb} compares the total HFB energy of the
fissioning nucleus just before and just after scission. In general,
scission is accompanied by a marked drop in HFB energy. That drop,
however, is much more pronounced for fission near the symmetric limit,
where it can be as large as $\sim50$ MeV over the $\Delta\left\langle \hat{Q}_{20}\right\rangle =5\,\textrm{b}$
change in quadrupole moment. Note that the fragment masses in Fig.
\ref{cap:240pu-frag-ehfb} are not the same before and after scission.
This difference is an indication of the drastic variations in the
nuclear density, and the redistribution of particles in the neck between
the two fragments at scission.

\begin{figure}
\includegraphics[%
  scale=0.33,
  angle=-90]{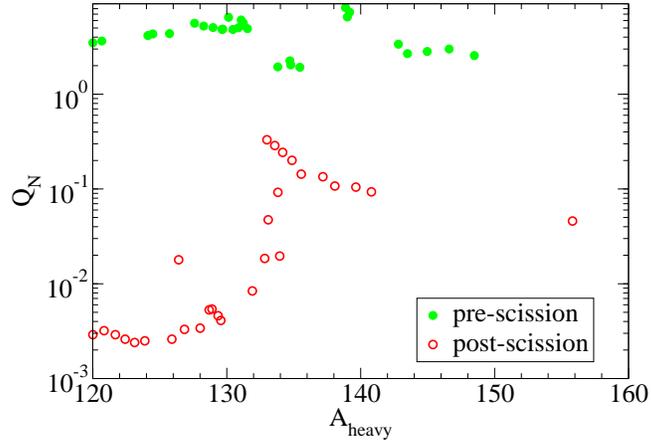}

\caption{\label{cap:240pu-frag-qn}(Color online) Number of particles in the
neck of the fissioning nucleus, plotted as a function of the heavy-fragment
mass number, obtained from the HFB calculations just before (solid
green disks) and just after (empty red circles) scission in Fig. \ref{cap:240pu-scl23}.}
\end{figure}

The number of particles in the neck just before and after scission
is shown in Fig. \ref{cap:240pu-frag-qn} as a function of the heavy-fragment
mass. The variation in $\left\langle \hat{Q}_{N}\right\rangle $ is
quite large (typically by an order of magnitude, but near the symmetric
limit, by more than a factor of 1000). 

\begin{figure}
\includegraphics[%
  scale=0.33,
  angle=-90]{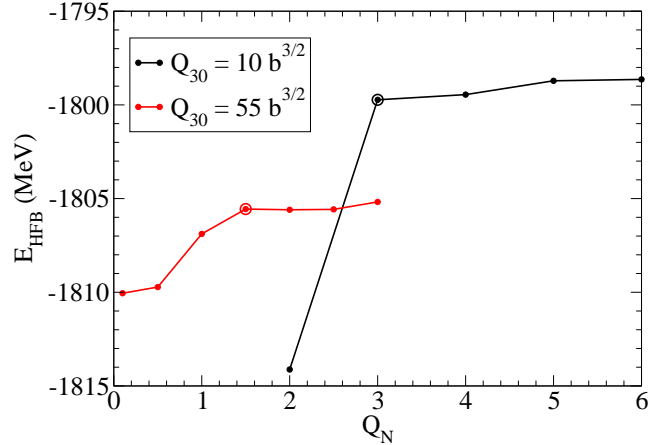}

\caption{\label{cap:breaking-qn}(Color online) Identification of the last
configuration before scission for HFB calculations at fixed $Q_{30}$=
$10\,\textrm{b}^{3/2}$ and $55\,\textrm{b}^{3/2}$, as a function
of the $Q_{N}$ constraint. The circled points on each curve were
chosen as the last pre-scission configuration, before the drop in
HFB energy as a function of decreasing $Q_{N}$.}
\end{figure}

\begin{figure}
\includegraphics[%
  scale=0.33,
  angle=-90]{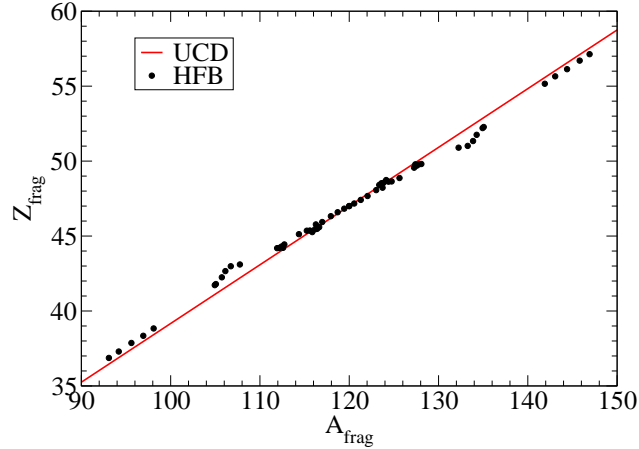}

\caption{\label{cap:240pu-frag-z}(Color online) Fission-fragment charge number
plotted as a function of mass number, obtained from the HFB calculations
immediately prior to scission in Fig. \ref{cap:240pu-scl23}. The
UCD prediction (solid red line) is plotted for comparison.}
\end{figure}

As in \cite{dubray08}, we extract the fragment properties for each
mass division from the HFB calculation just before scission. However,
we go further than the calculation in \cite{dubray08} by attempting
to approach the scission configuration even more closely. We introduce
an additional constraint on $Q_{N}$ to each point in the $Q_{20}-Q_{30}$
map of Fig. \ref{cap:240pu-scl23} just before the scission line,
and search for the $Q_{N}$ value marking a point just before a drop
in $E_{HFB}$ occurs. Fig. \ref{cap:breaking-qn} shows some typical
choices for this point. In Fig. \ref{cap:240pu-frag-z}, the charge
and mass of each fragment is plotted, covering a range from $A=93$
to $147$. We note that there is a nearly linear relationship between
the mass and charge of the fragments, which can be fitted as\begin{eqnarray*}
Z & = & 3.5349+0.36221\, A\end{eqnarray*}
This result is consistent with the prediction of the Unchanged-Charge
Division (UCD) model \cite{wahl88}, also shown in Fig. \ref{cap:240pu-frag-z}
for comparison, which for $^{240}\textrm{Pu}$ yields\begin{eqnarray*}
Z & = & \frac{94}{240}A\approx0.3917\, A\end{eqnarray*}

\begin{figure}
\includegraphics[%
  scale=0.33,
  angle=-90]{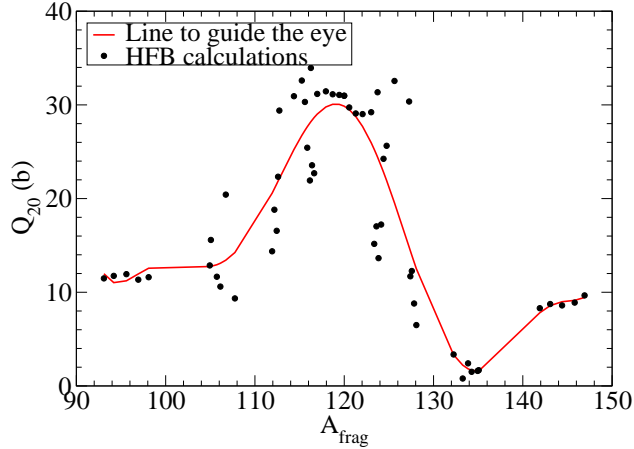}

\caption{\label{cap:240pu-frag-q20}(Color online) Fission-fragment quadrupole
moments, plotted as a function of fragment mass number, obtained from
the HFB calculations immediately prior to scission in Fig. \ref{cap:240pu-scl23}.
A line has been drawn through the HFB results to guide the eye.}
\end{figure}

\begin{figure}
\includegraphics[%
  scale=0.33,
  angle=-90]{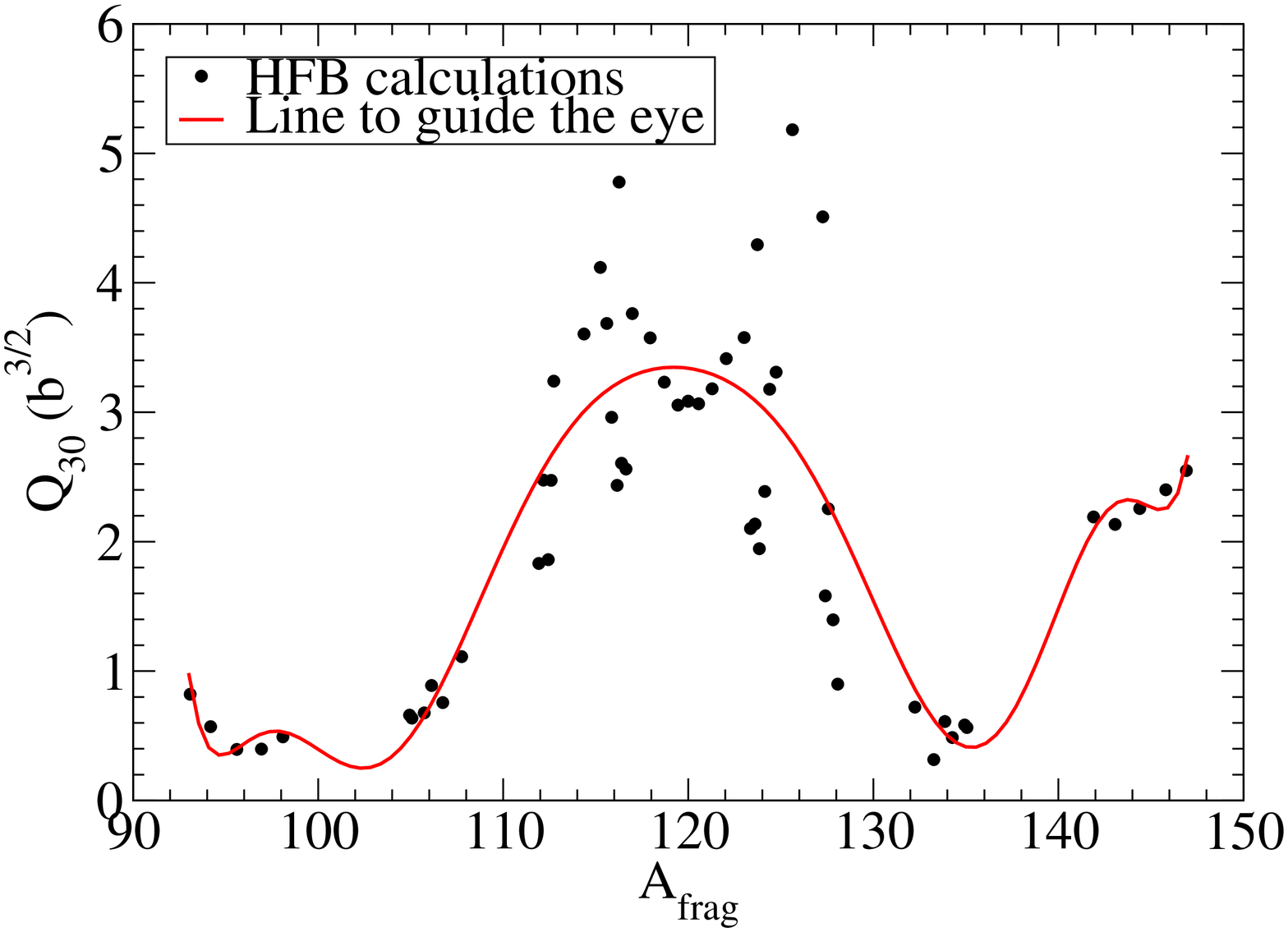}

\caption{\label{cap:240pu-frag-q30}(Color online) Same as Fig. \ref{cap:240pu-frag-q20},
but for the fission-fragment octupole moments.}
\end{figure}

\begin{figure}
\includegraphics[%
  scale=0.33,
  angle=-90]{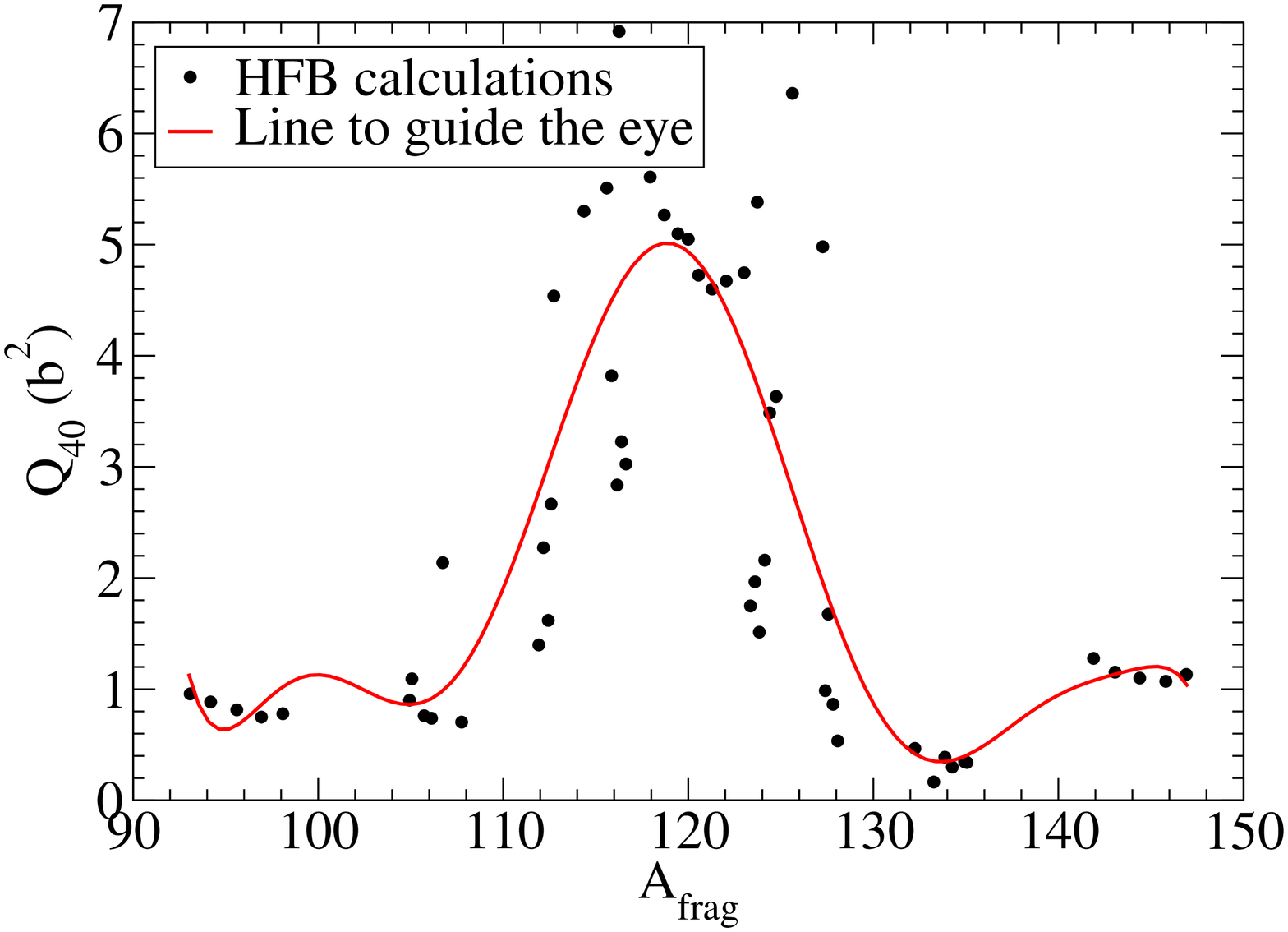}

\caption{\label{cap:240pu-frag-q40}(Color online) Same as Fig. \ref{cap:240pu-frag-q20},
but for the fission-fragment hexadecapole moments.}
\end{figure}

The moments of the fragments are shown in Figs. \ref{cap:240pu-frag-q20}-\ref{cap:240pu-frag-q40}.
The overall shape of the quadrupole moment in Fig. \ref{cap:240pu-frag-q20}
is similar to the one shown for $^{226}\textrm{Th}$ in Fig. \ref{cap:226th-a-q20},
with a maximum at the symmetric limit, and a drop-off on either side.
There is also a significant dip in the $\left\langle \hat{Q}_{20}\right\rangle $
value near the nearly-spherical $^{134}\textrm{Te}$ fragment. The
fragment octupole moment, plotted in Fig. \ref{cap:240pu-frag-q30},
also shows similarities in shape as well as magnitude to the $^{226}\textrm{Th}$
case in Fig. \ref{cap:226th-a-q30} %
\footnote{In our original HFB calculations for $^{240}\textrm{Pu}$, the $\left\langle \hat{Q}_{30}\right\rangle $
values for the light fragments are negative, but since the sign carries
no relevant physical meaning for this quantity, we have taken its
absolute value in Fig. \ref{cap:240pu-frag-q30}.%
}. Finally, we also show the hexadecapole moment of the fragments in
Fig. \ref{cap:240pu-frag-q40}. There as well, the value of $\left\langle \hat{Q}_{40}\right\rangle $
reaches a maximum near the symmetric limit, and drops off on either
side. In all cases, a line has been drawn to guide the eye using a
polynomial fit to the points. The HFB calculations in Figs. \ref{cap:240pu-frag-q20}-\ref{cap:240pu-frag-q40}
exhibit a great deal of fluctuation about the smooth polynomial fit.
These fluctuations are due for the most part to the difficulty in
identifying a scission configuration based on the criterion of sudden
changes in global nuclear properties, such as the total energy. In
a forthcoming paper \cite{younes09a}, we will embark on a more detailed
study of the scission configurations at the microscopic level, and
extract the excitation, kinetic, and interaction energies of the fragments.
The merits and difficulties of a scission criterion based on the interaction
energy between the fragments will be discussed in detail.

\section{Conclusion}

We have developed the HFB code FRANCHBRIE for microscopic fission
studies using the finite-range D1S effective interaction. The code
allows for the multiple constraints needed to explore the nuclear
densities relevant to fission, and is based on matrix elements calculated
in a one-center deformed harmonic-oscillator basis. We have provided
a detailed derivation of the formalism required for the adjustment
of those multiple constraints.

We have applied the code to the calculation of scission configurations
in the hot fission of $^{240}\textrm{Pu}$. These calculations are
relevant to studies of thermal neutron-induced fission on a target
of $^{239}\textrm{Pu}$. We have focused on the technical aspects
of using the HFB formalism for fission studies. In particular, we
have discussed some aspects of fission calculations within a one-center
basis, and the importance the choice of collective coordinates in
the HFB iterations for nearly-scissioned configurations. A scission
line in the quadrupole-octupole plane was obtained and shows a tendency
for the nucleus to reach much larger elongations in the symmetric
limit before scission occurs. A similar feature was observed in the
scission line of $^{226}\textrm{Th}$ by Dubray \emph{et al}. \cite{dubray08}
using two-center HFB calculations, reproduced in this work with a
one-center calculation. The increased {}``malleability'' of the
nucleus near the symmetric limit is reflected in the various moments
(quadrupole, octupole, hexadecapole) calculated for the fission fragments
and presented here.

In a forthcoming publication, we will extract the excitation and kinetic
energies of the fission fragments. We will introduce a microscopic
criterion for the identification of fission fragments, and calculate
their interaction energies, with special attention to the density
tails discussed in this paper. Finally, the static calculations of
hot fission presented here are the first step in a fully dynamical
calculation of $^{240}\textrm{Pu}$ fission. Further developments
are planned to explore all fission modes, from hot to cold, and to
include the dynamical aspects of the theory in the calculations.

\begin{acknowledgments}
This work was performed under the auspices of the U.S. Department
of Energy by the Lawrence Livermore National Laboratory under Contract
DE-AC52-07NA27344.
\end{acknowledgments}
\appendix

\section{Multiple constraint formalism\label{sec:Multiple-constraint-formalism}}

\subsection{Effect of the variation of a single Lagrange multiplier on the generalized
density}

In this appendix, we derive the formalism for solving the HFB equation
with multiple constraints. The derivation generalizes the discussion
in \cite{decharge80} to the case of multiple constraints. 

In the first section, we give the essential formulas used in the adjustment
of constraints. A second section illustrates the formalism with the
special case of a single constraint, and the last section presents
the general case of multiple constraints. Starting from the HFB equation,
Eq. (\ref{eq:HFB-equation}), we write for a Hamiltonian with a single
constraint $\lambda\hat{F}$ introduced as is Eq. (\ref{eq:constrd-ham}),\begin{eqnarray*}
\left[\mathcal{H}\left(R\left(\lambda\right),\lambda\right),R\left(\lambda\right)\right] & = & 0\end{eqnarray*}
where\begin{eqnarray}
\left\langle \lambda\left|\hat{F}\right|\lambda\right\rangle  & \equiv & f\left(\lambda\right)\nonumber \\
 & = & \frac{1}{2}\textrm{Tr}\hat{F}+\frac{1}{2}\textrm{Tr}\,\mathbb{F}R\left(\lambda\right)\label{eq:avgF-lam}\end{eqnarray}
is the expectation value of $\hat{F}$ in the corresponding HFB solution
$\left|\lambda\right\rangle $, with $\mathbb{F}$ given by Eq. (\ref{eq:F-mat-struct}).
Consider a small variation $\delta\lambda$ of the Lagrange multiplier,
leading to a new HFB solution with\begin{eqnarray}
\left[\mathcal{H}\left(R\left(\lambda+\delta\lambda\right),\lambda+\delta\lambda\right),R\left(\lambda+\delta\lambda\right)\right] & = & 0\label{eq:HFB-equation-pert}\end{eqnarray}
where\begin{eqnarray}
\left\langle \lambda+\delta\lambda\left|\hat{F}\right|\lambda+\delta\lambda\right\rangle  & \equiv & f\left(\lambda+\delta\lambda\right)\nonumber \\
 & = & \frac{1}{2}\textrm{Tr}\hat{F}\nonumber \\
 &  & +\frac{1}{2}\textrm{Tr}\,\mathbb{F}R\left(\lambda+\delta\lambda\right)\label{eq:avgF-lamdlam}\end{eqnarray}
We will now derive an explicit relation between the generalized density\begin{eqnarray*}
R\left(\lambda\right) & \equiv & ^{\left(0\right)}R\end{eqnarray*}
and its perturbed value, expanded to first order in $\delta\lambda$,\begin{eqnarray*}
R\left(\lambda+\delta\lambda\right) & \equiv & ^{\left(0\right)}R+{}^{\left(1\right)}R\end{eqnarray*}
Note that the idempotence condition in Eq. (\ref{eq:R-idempotence})
implies that the matrix ${}^{\left(1\right)}R$ has the form\begin{eqnarray}
^{\left(1\right)}\tilde{R} & = & \left(\begin{array}{cc}
0 & ^{\left(1\right)}\tilde{R}^{12}\\
^{\left(1\right)}\tilde{R}^{21} & 0\end{array}\right)\label{eq:1R-matrix-struct}\end{eqnarray}
in the quasiparticle representation that diagonalizes $^{\left(0\right)}R$.
A straightforward linearization of Eq. (\ref{eq:HFB-equation-pert})
about $^{\left(0\right)}R$ gives the relation\begin{eqnarray}
^{\left(1\right)}\vec{R} & = & \delta\lambda M^{-1}\vec{F}\label{eq:solve-for-1R}\end{eqnarray}
where $M$ is the QRPA matrix, whose elements are given by second-order
derivatives of the energy with respect to the generalized density
matrix \cite{decharge80}, and where we have introduced the vector
notation\begin{eqnarray}
\vec{F} & = & \left(\begin{array}{c}
F^{\left(1,2\right)}\\
F^{\left(1,2\right)*}\end{array}\right)\label{eq:MvecDef}\end{eqnarray}
and similarly for $^{\left(1\right)}\vec{R}$. Next, from Eqs. (\ref{eq:avgF-lam})
and (\ref{eq:avgF-lamdlam}), we deduce\begin{eqnarray}
\delta f & \equiv & f\left(\lambda+\delta\lambda\right)-f\left(\lambda\right)\nonumber \\
 & = & \frac{1}{2}\vec{F}^{\dagger}\cdot{}^{\left(1\right)}\vec{R}\label{eq:DF-def}\end{eqnarray}
Combining this result with Eq. (\ref{eq:solve-for-1R}), we can express
$\delta\lambda$ in the form\begin{eqnarray}
\delta\lambda & = & \frac{2\delta f}{\vec{F}^{\dagger}\cdot\left(M^{-1}\vec{F}\right)}\label{eq:dlambda-1constr}\end{eqnarray}
Equations (\ref{eq:solve-for-1R}) and (\ref{eq:dlambda-1constr})
are the basis for the iterative procedure described in the next section
that is used to solve the HFB equation under constraint.

In order to obtain a computationally efficient expression for the
inverse QRPA matrix $M^{-1}$ in Eq. (\ref{eq:dlambda-1constr}),
we adopt the so-called {}``cranking'' approximation where the residual
interaction between quasiparticles is neglected in the QRPA matrix.
In this case, $M^{-1}$ takes the block-diagonal form\begin{eqnarray*}
M^{-1} & = & \left(\begin{array}{cc}
\left[\left(\varepsilon_{\mu}+\varepsilon_{\nu}\right)^{-1}\delta_{\mu\sigma}\delta_{\nu\tau}\right] & \left[0\right]\\
\left[0\right] & \left[\left(\varepsilon_{\mu}+\varepsilon_{\nu}\right)^{-1}\delta_{\mu\sigma}\delta_{\nu\tau}\right]\end{array}\right)\end{eqnarray*}
and therefore,\begin{eqnarray}
{}^{\left(1\right)}R_{\mu\nu}^{21} & = & \frac{\delta\lambda}{\varepsilon_{\mu}+\varepsilon_{\nu}}\sum_{mn}\left(F_{mn}V_{m\mu}U_{n\nu}\right.\nonumber \\
 &  & \left.-F_{mn}^{*}U_{m\mu}V_{n\nu}\right)\label{eq:R1-cranking}\end{eqnarray}
with a corresponding expression for $\delta\lambda$.

\subsection{Adjustment of the HFB solution in the case of one constraint\label{sub:single-adjustment}}

In this section, we examine in greater detail steps \ref{enu:Calculate-deltalambdai-needed}
and \ref{enu:Calculate-deltaR} in the description of the HFB algorithm
listed in section \ref{sub:The-HFB-convergence}. In this case, the
constrained HFB equation is written\begin{eqnarray*}
\left[\mathcal{H}\left(R\right)-\lambda\mathbb{F},R\right] & = & 0\end{eqnarray*}
with\begin{eqnarray*}
f & = & \frac{1}{2}\textrm{Tr}\hat{F}+\frac{1}{2}\textrm{Tr}\,\mathbb{F}R\end{eqnarray*}
where $f$ is the expectation value of the constraint operator. The
solution of the HFB equation then consists not only in determining
$R$, but also the Lagrange multiplier $\lambda$ that satisfies the
constraint. To solve this problem, we are led to an iterative procedure
wherein the Lagrange multiplier is adjusted at each iteration. Consider
the $n^{\textrm{th}}$ iteration, such that the generalized density
matrix obtained in the previous iteration is $R^{\left(n-1\right)}$
with a corresponding Lagrange multiplier $\lambda^{\left(n-1\right)}$.
The diagonalization of $\mathcal{H}\left(R^{\left(n-1\right)}\right)-\lambda^{\left(n-1\right)}\mathbb{F}$
leads to a new generalized density which we will denote $\bar{R}^{\left(n\right)}$.
At this stage, the constraint is no longer necessarily satisfied and
we calculate the deviation from the desired value\begin{eqnarray*}
\delta f^{\left(n\right)} & = & f-f^{\left(n\right)}\end{eqnarray*}
We correct the Lagrange multiplier using Eq. (\ref{eq:dlambda-1constr}),\begin{eqnarray*}
\lambda^{\left(n\right)} & = & \lambda^{\left(n-1\right)}+\frac{2\delta f^{\left(n\right)}}{\vec{F}^{\dagger}\cdot\left(M^{-1}\vec{F}\right)}\end{eqnarray*}
and the generalized density using Eq. (\ref{eq:solve-for-1R}),\begin{eqnarray*}
R^{\left(n\right)} & = & \bar{R}^{\left(n\right)}+\delta\lambda M^{-1}\vec{F}\end{eqnarray*}
with\begin{eqnarray*}
\delta\lambda & = & \lambda^{\left(n\right)}-\lambda^{\left(n-1\right)}\end{eqnarray*}
We define the $n^{\textrm{th}}$ iteration with the self-consistent
pair of $R^{\left(n\right)}$ and $\lambda^{\left(n\right)}$. Note
that the constraint is satisfied at each iteration. This iterative
process generally converges, i.e.\begin{eqnarray*}
R^{\left(n\right)} & \rightarrow & \bar{R}^{\left(n\right)}\rightarrow R\\
\lambda^{\left(n\right)} & \rightarrow & \lambda\\
f^{\left(n\right)} & \rightarrow & f\end{eqnarray*}
If the difference in constraint values is very large between successive
iterations (as may be the case in the first few iterations), the convergence
rate can be improved by calculating the generalized density matrix
at the $n^{\textrm{th}}$ iteration according to\begin{eqnarray*}
R^{\left(n\right)} & = & \left(1-\alpha\right)\left(\bar{R}^{\left(n\right)}+\delta\lambda M^{-1}\vec{F}\right)+\alpha R^{\left(n-1\right)}\end{eqnarray*}
with the associated Lagrange multiplier\begin{eqnarray*}
\lambda^{\left(n\right)} & = & \left(1-\alpha\right)\left(\lambda^{\left(n-1\right)}+\delta\lambda\right)+\alpha\lambda^{\left(n-1\right)}\end{eqnarray*}
where the weight $\alpha$ tends to zero as the solution converges.
With this prescription, the convergence of the generalized density
and Lagrange multiplier are slowed down by the same amount. In other
words, the desired value $f$ for the constraint is approached in
a gradual manner, so that at the $n^{\textrm{th}}$ iteration\begin{eqnarray*}
\left\langle \lambda^{\left(n\right)}\left|\hat{F}\right|\lambda^{\left(n\right)}\right\rangle  & = & f^{\left(n\right)}=\left(1-\alpha\right)f+\alpha f^{\left(n-1\right)}\end{eqnarray*}

\subsection{Adjustment of the HFB solution in the case of multiple constraints}

The results in the previous section can be readily generalized to
an arbitrary number $N$ of constraints. In this case, the HFB procedure
minimizes the energy\[
\left\langle \left\{ \lambda\right\} \left|H-\sum_{i=1}^{N}\lambda_{i}\hat{F}_{i}\right|\left\{ \lambda\right\} \right\rangle \]
subject to the set of constraints\begin{eqnarray*}
\left\langle \left\{ \lambda\right\} \left|\hat{F}_{i}\right|\left\{ \lambda\right\} \right\rangle  & = & f_{i},\quad i=1,\ldots,N\end{eqnarray*}
The generalized density matrix is now a function of $N$ Lagrange
multipliers, $R\left(\left\{ \lambda\right\} \right)$. We write\begin{eqnarray}
R\left(\left\{ \lambda+\delta\lambda\right\} \right)-R\left(\left\{ \lambda\right\} \right) & \equiv & ^{\left(1\right)}R\nonumber \\
 & = & \sum_{i=1}^{N}\frac{\partial R}{\partial\lambda_{i}}\delta\lambda_{i}\nonumber \\
 & = & \sum_{i=1}^{N}{}^{\left(1\right)}R_{i}\label{eq:1R-many}\end{eqnarray}
Clearly, $^{\left(1\right)}R_{i}$ is a variation where all the Lagrange
multipliers are held fixed except for the one associated with $\hat{F}_{i}$.
Therefore, $^{\left(1\right)}R_{i}$ is given by Eq. (\ref{eq:solve-for-1R})
with the substitutions $\delta\lambda\rightarrow\delta\lambda_{i}$
and $\hat{F}\rightarrow\hat{F}_{i}$. In the case of multiple constraints,
Eq. (\ref{eq:solve-for-1R}) is therefore replaced by\begin{eqnarray}
^{\left(1\right)}\vec{R} & = & \sum_{i=1}^{N}\delta\lambda_{i}M^{-1}\vec{F}_{i}\label{eq:solve-for-1R-many}\end{eqnarray}
Furthermore, using the generalization of Eq. (\ref{eq:DF-def}) to
multiple constraints,\begin{eqnarray*}
\delta f_{i} & \equiv & \left\langle \left\{ \lambda+\delta\lambda\right\} \left|\hat{F}_{i}\right|\left\{ \lambda+\delta\lambda\right\} \right\rangle -\left\langle \left\{ \lambda\right\} \left|\hat{F}_{i}\right|\left\{ \lambda\right\} \right\rangle \\
 & = & \frac{1}{2}\vec{F}_{i}^{\dagger}\cdot{}^{\left(1\right)}\vec{R}\end{eqnarray*}
and taking into account Eq. (\ref{eq:solve-for-1R-many}), we finally
obtain\begin{eqnarray}
\delta\lambda & = & T^{-1}\delta f\label{eq:dlambda-many-constr}\end{eqnarray}
where the $N\times N$ matrix $T$ is defined by\begin{eqnarray}
T_{lm} & \equiv & \frac{1}{2}\vec{F}_{l}^{\dagger}\cdot\left(M^{-1}\vec{F}_{m}\right)\label{eq:T-matrix-def}\end{eqnarray}
Note that this matrix introduces correlations between all the constraints.
We assume in our discussion that the inverse matrix $T^{-1}$ exists,
i.e., that the constraints are independent. Eqs. (\ref{eq:solve-for-1R-many})
and (\ref{eq:dlambda-many-constr}) then replace Eqs. (\ref{eq:solve-for-1R})
and (\ref{eq:dlambda-1constr}) in the adjustment method described
above.

\section{Translation in a finite harmonic oscillator basis\label{sec:Translation-in-a}}

In this section, we give the explicit form for the expansion of a
translated harmonic-oscillator function in a harmonic-oscillator basis.
We begin with the generating function for the Cartesian harmonic-oscillator
function (Eq. (A.1) in \cite{younes09}).\begin{eqnarray}
e^{-t^{2}+2tx/b-x^{2}/\left(2b^{2}\right)} & = & \sqrt{b\sqrt{\pi}}\sum_{k=0}^{\infty}\frac{2^{k/2}}{\sqrt{k!}}t^{k}\Phi_{k}\left(x;b\right)\label{eq:genfun-ho-zfun}\end{eqnarray}
Letting $x\rightarrow x+\Delta x$ on both sides of Eq. (\ref{eq:genfun-ho-zfun})
after some simplification, the left-hand side (LHS) can be written
as\begin{eqnarray*}
LHS & = & \sqrt{b\sqrt{\pi}}e^{-\Delta x\left(x+\Delta x/2\right)/b^{2}}\\
 &  & \times\sum_{m=0}^{\infty}\sum_{n=0}^{\infty}\frac{2^{m+n/2}\left(\Delta x/b\right)^{m}}{m!\sqrt{n!}}\Phi_{n}\left(x;b\right)t^{m+n}\end{eqnarray*}
where we have used Eq. (\ref{eq:genfun-ho-zfun}) to express the LHS
in terms of harmonic-oscillator functions. Equating like powers of
the arbitrary variable $t$ between the LHS and right-hand side (RHS),
we obtain\begin{eqnarray}
\Phi_{k}\left(x+\Delta x;b\right) & = & e^{-\Delta x\left(x+\Delta x/2\right)/b^{2}}\nonumber \\
 &  & \times\sum_{m=0}^{k}\frac{2^{m/2}\sqrt{k!}\left(\Delta x/b\right)^{m}}{m!\sqrt{\left(k-m\right)!}}\nonumber \\
 &  & \times\Phi_{k-m}\left(x;b\right)\label{eq:phi-disp1}\end{eqnarray}
This is still a finite sum over harmonic-oscillator functions, however
an overall exponential factor depending on $x$ remains, and must
be eliminated in order to obtain the expansion of $\Phi_{k}\left(x+\Delta x;b\right)$
on the harmonic-oscillator basis. Thus, in general, we need to derive
an expansion for the expression\begin{equation}
e^{2\alpha x/b^{2}}\Phi_{i}\left(x;b\right)\label{eq:expfac-times-phi}\end{equation}
where $\alpha=-\Delta x/2$ and $i=k-m$ in our case. Starting from
the generating function in Eq. (\ref{eq:genfun-ho-zfun}), and multiplying
both sides by the exponential factor in Eq. (\ref{eq:expfac-times-phi}),
the LHS of Eq. (\ref{eq:genfun-ho-zfun}) becomes after some simplification\begin{eqnarray*}
LHS & = & \sqrt{b\sqrt{\pi}}e^{\alpha^{2}/b^{2}}\sum_{l=0}^{\infty}\frac{2^{l/2}}{\sqrt{l!}}e^{2\alpha t/b}\left(t+\frac{\alpha}{b}\right)^{l}\Phi_{l}\left(x;b\right)\end{eqnarray*}
Expanding in powers of the arbitrary variable $t$, this takes the
form\begin{eqnarray*}
LHS & = & \sqrt{b\sqrt{\pi}}e^{\alpha^{2}/b^{2}}\sum_{l=0}^{\infty}\sum_{p=0}^{\infty}\sum_{q=0}^{l}\\
 &  & \times\left(\begin{array}{c}
l\\
q\end{array}\right)\frac{2^{p+l/2}}{p!\sqrt{l!}}\left(\frac{\alpha}{b}\right)^{l+p-q}\Phi_{l}\left(x;b\right)t^{p+q}\end{eqnarray*}
Therefore, equating like powers of $t$ between LHS and RHS, we obtain\begin{eqnarray*}
e^{2\alpha x/b^{2}}\Phi_{k}\left(x;b\right) & = & e^{\alpha^{2}/b^{2}}\sum_{l=0}^{\infty}\sum_{q=0}^{l}\left(\begin{array}{c}
l\\
q\end{array}\right)\frac{2^{\left(k+l\right)/2-q}\sqrt{k!}}{\left(k-q\right)!\sqrt{l!}}\\
 &  & \times\left(\frac{\alpha}{b}\right)^{l+k-2q}\Phi_{l}\left(x;b\right)\end{eqnarray*}
Using this result in Eq. (\ref{eq:phi-disp1}), we obtain\begin{eqnarray}
\Phi_{k}\left(x+\Delta x;b\right) & = & e^{-\Delta x^{2}/\left(4b^{2}\right)}\nonumber \\
 &  & \times\sum_{l=0}^{\infty}C_{l}\left(-\frac{\Delta x}{2b}\right)\Phi_{l}\left(x;b\right)\label{eq:phi-disp2}\end{eqnarray}
where\begin{eqnarray*}
C_{l}\left(\xi\right) & = & 2^{\left(k+l\right)/2}\sqrt{\frac{k!}{l!}}\xi^{\left(k+l\right)/2}\\
 &  & \times\sum_{m=0}^{k}\sum_{q=0}^{l}\frac{\left(-1\right)^{m}2^{m-q}}{m!\left(k-m-q\right)!}\left(\begin{array}{c}
l\\
q\end{array}\right)\xi^{-2q}\end{eqnarray*}
Note that the expansion of the translated harmonic-oscillator function
requires in principle and infinite number of terms. In practice, these
translations are performed in a finite-sized basis, and the truncation
of the sum in Eq. (\ref{eq:phi-disp2}) to those shells within the
basis can lead to the appearance of tails for translated nuclear densities
expanded in a finite harmonic-oscillator basis, as shown in Figs.
\ref{cap:transfrag} and \ref{cap:transfrag1}.

\end{document}